\documentclass[a4paper, reprint, aps,prd,longbibliography, twoside]{revtex4-1}

\usepackage[english]{babel}
\usepackage[utf8]{inputenc}
\usepackage[x11names]{xcolor}
\usepackage{float,comment}
\usepackage[colorinlistoftodos, color=green!40, prependcaption]{todonotes}
\usepackage{hyperref}
\usepackage{mathtools}
\usepackage[top=1.in, bottom=1.in, left=0.5in, right=0.5in]{geometry}
\usepackage{amsthm, amssymb}
\usepackage{graphicx}
\usepackage{lipsum}
\usepackage{svg}
\usepackage{pbox}
\usepackage{color}

\DeclarePairedDelimiter\ton{(}{)}
\DeclarePairedDelimiter\qua{[}{]}

\DeclarePairedDelimiter\mean{<}{>}
\newcommand{\me}{\mathrm{e}}

\begin{document}
\title{Dynamical approach to Zipf's law}

\author{Giordano De Marzo$^1$}
\author{Andrea Gabrielli$^{2,3}$}
\author{Andrea Zaccaria$^3$}
\email[Correspondence email address: ]{andrea.zaccaria@cnr.it}
\author{Luciano Pietronero$^{1,3, 4}$}
    
    \affiliation{$^1$Dipartimento di Fisica Universit\`a ``Sapienza”, P.le A. Moro, 2, I-00185 Rome, Italy.\\
    $^2$Dipartimento di Ingegneria, Universit\`a Roma 3, Via Vito Volterra 62, I-00146 Rome, Italy\\
    $^3$Istituto dei Sistemi Complessi (ISC) - CNR, UoS Sapienza,P.le A. Moro, 2, I-00185 Rome, Italy.\\
    $^4$Centro Ricerche Enrico Fermi, Piazza del Viminale, 1,
I-00184 Rome, Italy}

\date{\today} 

\begin{abstract}
The rank-size plots of a large number of different physical and
socio-economic systems are usually said to follow Zipf’s law, but a
unique framework for the comprehension of this ubiquitous scaling law
is still lacking. Here we show that a dynamical approach is crucial:
during their evolution, some systems are attracted towards Zipf’s law,
while others presents Zipf’s law only temporarily and, therefore,
spuriously. A truly Zipfian dynamics is characterized by a dynamical constraint, or coherence, among the parameters of the generating PDF, and the number of elements in the system. A clear-cut example of such coherence is natural language. Our framework allows us to derive some quantitative results that go well beyond the usual Zipf's law: i) earthquakes can evolve only incoherently and thus show Zipf’s law spuriously; this allows an
assessment of the largest possible magnitude of an earthquake
occurring in a geographical region. ii) We prove that Zipfian dynamics are not additive, explaining analytically why US cities evolve coherently, while world cities do not. iii) Our concept of coherence can be used for model selection, for example, the Yule-Simon process can describe the dynamics of world countries' GDP. iv) World cities present spurious Zipf's law and we use this property for estimating the maximal population of an urban agglomeration. 
\end{abstract}


\maketitle

 \section{Introduction}
Zipf's law \cite{Language, Jaguar} is an empirical scaling relation that connects the sizes of a set of objects with their ranking when sorted according to the size itself. Being ubiquitous in nature, it represents one of the most studied topics in complex systems: it has been observed in the size distribution of cities \cite{Batty}, of firms \cite{Firms} and of GDPs \cite{Batty}, but also in natural language \cite{Language}, in web page visits \cite{Web}, in scientific citations \cite{Citations, Newman} and many natural systems, such as earthquakes \cite{Newman, sornette1996rank} and lunar craters \cite{Newman}. 
There are currently numerous approaches to explain Zipf's law based on different mechanisms including multiplicative processes \cite{Multiplicative, levy1996power}, adjacent possible framework \cite{Loreto, Kauffman}, sample space reducing processes \cite{SSR}, and information theory arguments \cite{Marsili, Mandelbrot}. However, while all these models give insights on the upset of Zipf's scaling, they fail in providing a general explanation of the phenomenon. \\

In this work we show that the dynamical evolution in space or time of these models and natural systems, when analyzed from the perspective of Zipf's law, provides unprecedented quantitative insights.
A first result is that while some systems are dynamically attracted toward Zipf's law, others show Zipf's law only temporarily and, as a consequence, we can label them as \textit{spuriously} Zipfian. This relatively simple observation has a number of crucial implications. Indeed, \textit{spuriously} Zipfian systems during their evolution deviate from Zipf's law by sampling the tail of the (power-law) probability density function (PDF) from which their sizes are extracted. More precisely large events are completely sampled. This allows to determine the upper cutoff of the PDF and since we demonstrate that earthquakes follow Zipf's law only spuriously, we can determine the maximum possible magnitude of an earthquake occurring in a given geographical region. Conversely, systems for which Zipf's law is a dynamical attractor are intrinsically under-sampled and, as such, in a sort of permanent out of equilibrium or transient state. These \textit{genuine} Zipfian systems are characterized by a dynamical constraint, that we call \textit{coherence}, relating the sampling rate to the lower and upper cutoffs of the generating PDF.\\
Natural language, the first and most famous application of Zipf's law, naturally satisfies such constraint due to the presence of grammar and semantic rules. Moreover, our approach allows to understand why Zipf's law holds for the cities of single countries, but not for larger sets of nations, a phenomenon abundantly debated in the literature \cite{Batty}. Finally, we applied our dynamical approach also to generative models such as multiplicative processes \cite{levy1996power} and the Yule-Simon model \cite{yule1925ii, simon1955class}. We analytically show that both processes are genuinely Zipfian and, as a by-product, that the preferential attachment mechanism reproduces the dynamical evolution of world countries. \\
Using these results we can provide also a novel insight about Heaps' law \cite{heaps1978information}, a likewise ubiquitous scaling law whose connection with Zipf's law has been largely debated. By generalizing previous analyses \cite{lu2010}, we demonstrate that Heapsian systems are a subset of the Zipfian ones. This allows us to predict that US cities evolved according to Heaps' scaling, a result confirmed by empirical analysis.\\
From this body of analysis it clearly emerges that our framework is completely general and can be applied to any system or model claimed to show Zipf's law. 


\section{Analytical results} 
\begin{figure*}[t]
\centering
\includegraphics[width=1\linewidth]{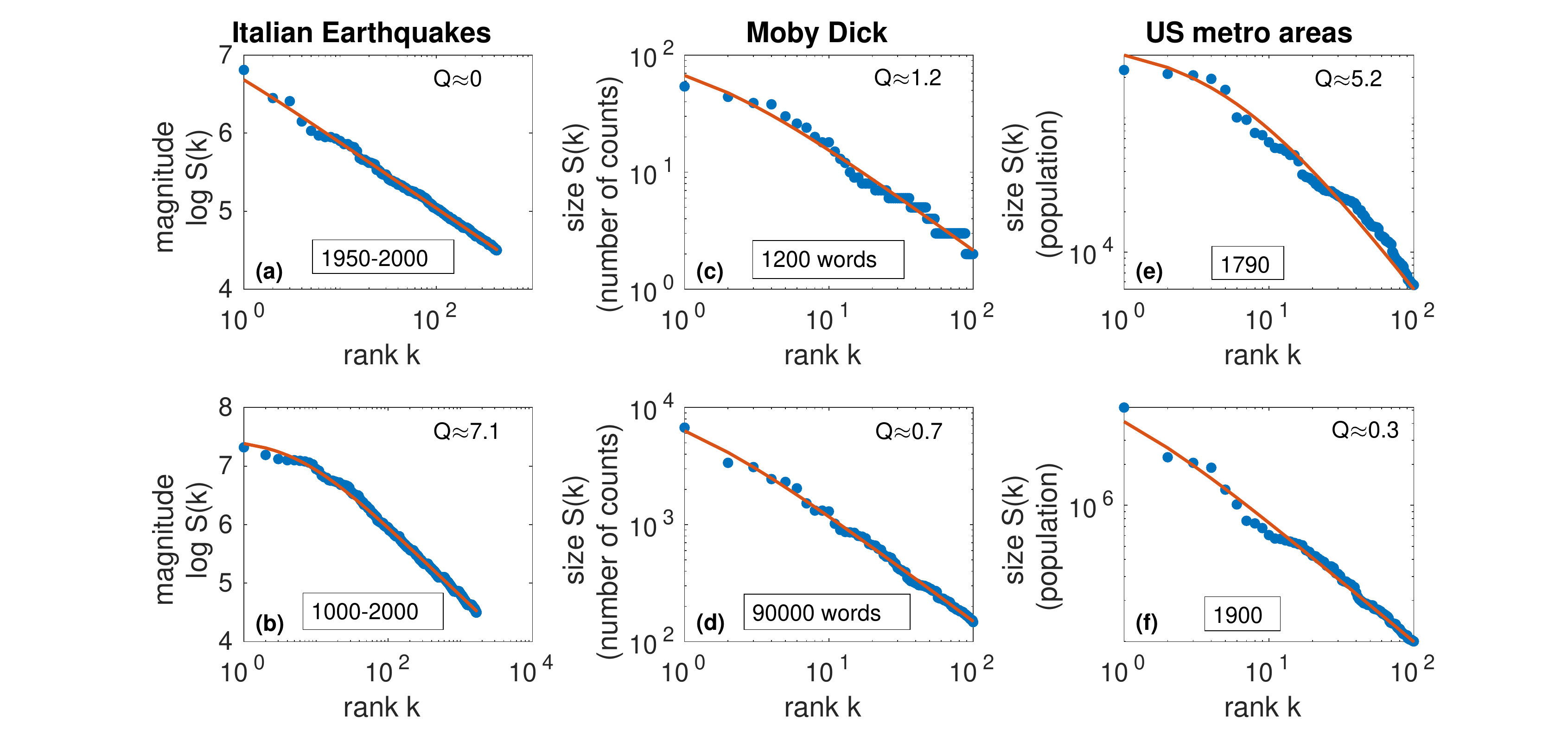}
\caption{\textbf{The deviation Q from a pure Zipf's law depends on the sampling.} (a) and (b): Rank-size plots of Italian earthquakes in the period 1900-2000 and 1000-2000. (c) and (d): rank-size plots using the first 1600 or the first 90000 words of the novel Moby Dick. (e) and (f): Rank-size plots of US metropolitan areas in 1790 and in 1900. For earthquakes the magnitude is the logarithm of the size, for metro areas the size is the population, while in the case of words the size is the number of occurrences. In all three systems, a different sampling leads to a more or less Zipfian behaviour.}
\label{fig:earthquakes_MobyDick_Cities}
\end{figure*}

\subsection{Dynamical constraint determine Zipfian dynamics}
In order to fix the notation, we give a formal definition of Zipf's law. Given a set of $N$ objects and denoting by $S(k)$ the size of the $k$th largest one Zipf's law reads
\[
    S(k)=\frac{S(1)}{k^{\gamma}}.
\]
For instance, $N$ could be regarded as the number of cities in a country and $S(k)$ as the population of the $k$th most populous urban settlement. More generally, $S(k)$ can be fitted using the Zipf-Mandelbrot relation \cite{Mandelbrot}, which accounts for the presence of deviations at low ranks:
\begin{equation}
    S(k)=\frac{\bar{S}}{(k+Q)^{\gamma}},
    \label{eq:Zipf_Mandelbrot}
\end{equation}
where $\bar{S}$ and $Q$ are generally regarded as free parameters. Zipf's law is recovered for $\bar{S}=S(1)$ and $Q=0$: as such, $Q$ can be regarded as an empirical measure of the deviation from a pure Zipf's law, whose importance is also given by the specific involvement of the largest objects.\\
Zipf's law is usually associated to a power-law probability distribution function (PDF) according to which sizes are distributed \cite{Batty, Newman}. Real power laws are always characterized by an upper and a lower cutoff, which correspond to intrinsic physical limits and that will play a fundamental role in determining the Zipfian behavior of the system\footnote{The effect of the only upper cutoffs has been previously considered \cite{burroughs2001, burroughs2001b}; in any case, never from a dynamical point of view.}. Let us suppose to consider $N$ objects whose sizes are distributed according to a truncated power law PDF of the form 
		\begin{equation}
		    	P(S)=
			\begin{cases}
				0 \ \text{for} \ S<s_m \\
				\frac{c}{S^{\alpha}} \ \text{for} \ s_m\leq S \leq s_M \\
				0 \ \text{for} \ S>s_M
			\end{cases}
		\label{trunc}
		\end{equation}
where $s_m$ is the lower cutoff and $s_M$ the upper one. As we show in the Methods section, the parameters of the Zipf-Mandelbrot relation \eqref{eq:Zipf_Mandelbrot} describing these objects can be written as simple functions of $N$ and the parameters of the PDF:
		\begin{equation}
    	    \begin{cases}
    	        \gamma=\frac{1}{\alpha-1}\\
    		    \bar{S}=N^{\gamma}s_m\\
    		    Q=N\ton*{\frac{s_m}{s_M}}^{\frac{1}{\gamma}}\,.
    	    \end{cases}
    	    \label{eq:bar_s_Q}
        \end{equation}
Note that we also recovered the well known result $\gamma=\frac{1}{\alpha-1}$ which links the exponent of the PDF to the asymptotic exponent of the rank-size plot \cite{Batty, li2002}. Moreover $Q$ is explicitly related the level of sampling of the distribution, relating the extension of the PDF to the number of elements in the system, and it is a non-negative parameter. Substituting these expressions into the Zipf-Mandelbrot scaling law, we get the rank-size relation followed by $N$ objects whose sizes are power law distributed:
        \begin{equation}
    	    S(k)=\frac{N^{\gamma}s_m}{\qua*{k+N\ton*{\frac{s_m}{s_M}}^{\alpha-1}}^{\frac{1}{\alpha-1}}}\,.
    	    \label{eq:rank_size}
        \end{equation}

\begin{figure*}
		\centering
	 	\includegraphics[width=1\linewidth]{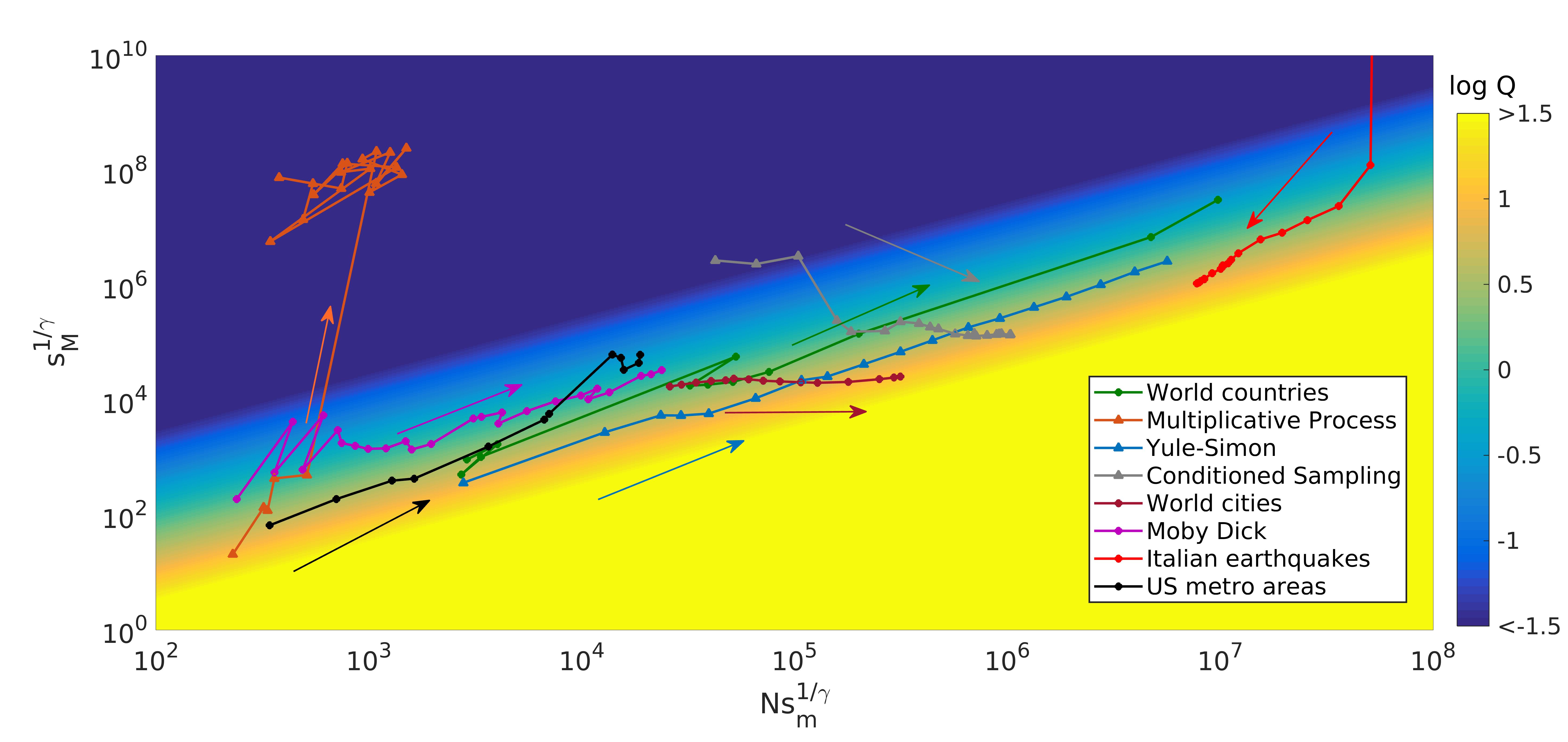}
\caption{\textbf{Zipf's plane.} Trajectories in Zipf's plane of US metropolitan areas, world cities, Italian earthquakes, world countries, the novel Moby Dick, a multiplicative process, the conditioned sampling and a typical trajectory of the Yule-Simon process. Real systems are denoted by circles, while models by triangles. We recall that for earthquakes the magnitude is the logarithm (in base $10$) of the size i.e. of the maximal amplitude detected, for cities the size is the population, in the case of words size is the number of occurrences while for countries the size is defined as the GDP {\em per capita}. Italian earthquakes and world cities are moved toward high values of $Q$ by dynamics and so evolve incoherently. Differently, Moby Dick, US metropolitan areas and World countries tend to the low $Q$ region and consequently show a Zipfian dynamics. For what concerns models, the conditioned sampling is not Zipfian, while the Yule-Simon model and the multiplicative process show Zipfian dynamics. However while the multiplicative process performs a trajectory very far from the ones of the real systems we considered, the evolution of the Yule-Simon process (blue triangles) is very similarly to the one of world countries (green circles).}
		\label{fig:zipf_plane}
	\end{figure*}	

All the systems where Zipf's law is found are, in a certain sense, dynamical: both the number $N$ of objects in the system and the parameters of the generating PDF will in general vary over time. Since the adherence to Zipf's law, as measured by $Q$, is a function of both $N$ and these parameters, it is then natural to investigate how $Q$ is influenced by the dynamics, so to determine if Zipf's law is \textit{dynamically stable}. Consider, for example, a very simple situation in which new objects are drawn with the passage of time, keeping the parameters of the PDF fixed. From Eqs.~\eqref{eq:bar_s_Q} it is clear that in this case the deviation parameter $Q$ is expected to increase with time. As a consequence, a static approach would address the system as Zipfian while $N$ remains relatively small, and would address it as not-Zipfian when $N$ becomes large with respect to $\ton*{\frac{s_M}{s_m}}^{\alpha-1}$. Clearly, one should consider the first situation as a \textit{spurious} manifestation of Zipf's law, because it is only due to a temporary under-sampling of the PDF. In other words, in this case Zipf's law is not stable and consequently it does not represent an attractor of the dynamics. 
			
Real systems are characterized by different and non trivial behaviors concerning the dynamics of the rank-size plot. Fig.~\ref{fig:earthquakes_MobyDick_Cities} provides some empirical examples of such dynamics. In the first column we show the rank-size plots of the earthquakes occurred in Italy, collected from INGV historical dataset \cite{Earthquakes}. We recall that in this case the size $S$ is defined as the exponential of the (moment) magnitude of the earthquake considered. In panel \textbf{(a)} we consider only the earthquakes occurred between 1900 and 2000, that perfectly adhere to Zipf's law, while in panel \textbf{(b)} those which occurred in the same area, but during a wider time window (1000-2000). In this last case the first ranks clearly deviate from a pure Zipfian scaling being $Q$ relatively large. Note, however, that other systems show a different trend when $N$ is increased, as in the case of the novel Moby Dick, second column of Fig.~\ref{fig:earthquakes_MobyDick_Cities}, and US metropolitan areas, third column of the same figure. Here the sizes are, respectively, the number of word occurrences and population. The population of US metro areas comes from the work of Schroeder \cite{Metro_areas}. In the case of Moby Dick, an increase of the words considered does not result in a increase of $Q$, which, instead, slightly decreases. The dynamics of US metro areas is even more peculiar and evident. Indeed metro areas were not distributed according to Zipf's law in 1790 (panel \textbf{(e)}), soon after the Declaration of Independence, while this scaling relation is found considering the same system at the beginning of the last century (panel \textbf{(f)}). In both these systems $N$ is increasing and the parameters of the PDF are varying, but, unlike from what happens considering earthquakes, they are varying coherently, that is, in such a way to make $Q$ decrease with the dynamics. The essence of a genuine Zipfian system, as opposed to a simpler set of objects whose sizes are drawn from a power law distribution, is then contained in the dynamical relations among $N$ and the parameters of the PDF $\alpha$, $s_m$ and $s_M$.

\subsection{Coherence of the  ``time" evolution}
\begin{figure*}
				\centering
			 	\includegraphics[width=1\linewidth]{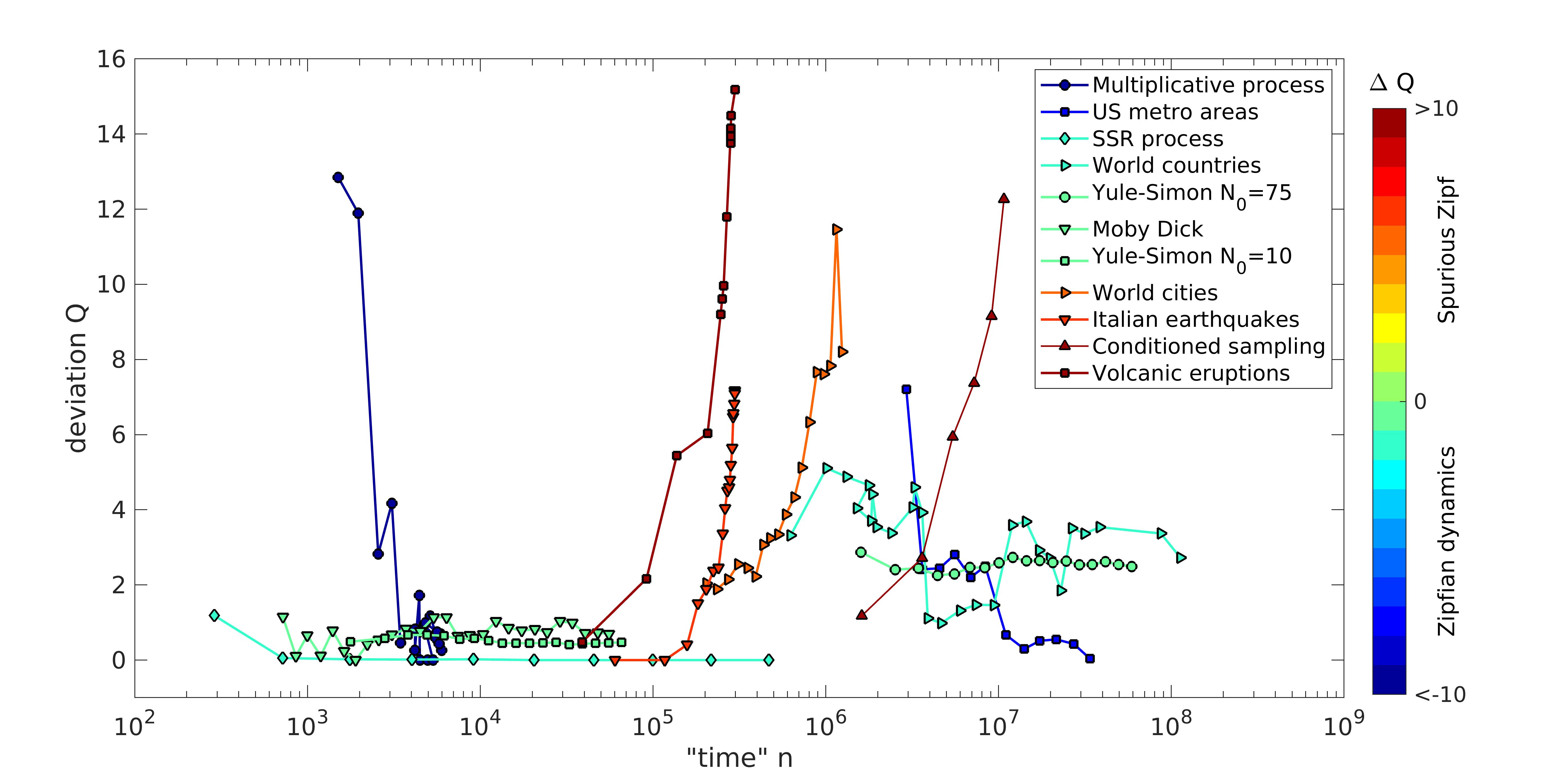}
		\caption{\textbf{The dynamics of deviations.} Evolution of the deviations parameter $Q$ as function of the ''time`` variable $n$ for different systems and models (see main text and Fig.~\ref{fig:zipf_plane}). Red trajectories are indicative of a spurious Zipf's law, for instance earthquakes and the conditioned sampling show this behavior. Differently, blue trajectories correspond to a Zipfian dynamics, and so to a genuine Zipf's law. Yule-Simon model and US metro areas are examples of this kind of dynamics. We considered also casualties provoked by volcanic eruptions \cite{noaaVolcanoes} and a typical trajectory of the Sample Space Reducing process \cite{SSR}.}
				\label{fig:n_Q}
			\end{figure*}

The considerations of the previous paragraph strongly suggest that an indicative study of Zipf's law can be performed only \textit{dynamically}, by checking the deviation parameter $Q$ as a function of time. Therefore, we propose to focus the attention not on the static identification of systems which show Zipf's law, but on the dynamics which makes $N$ and the parameters of the PDF evolve coherently, that is, in such a way that $Q$ decreases. We call this behavior \textbf{Zipfian dynamics}:
\begin{itemize}
	\item a system shows Zipfian dynamics when the rank-size plot, following the dynamics, does not present increasing deviations from a straight line. In mathematical terms, this is equivalent to requiring the underlying PDF to be a power law and the parameter $Q(n)$ to be not increasing with $n$, where $n$ is a variable which plays the role of time. 
\end{itemize}	
We call \textit{genuine} Zipfian a system which shows Zipfian dynamics.

In order to visualize the evolution of systems under this perspective, we introduce the \textit{Zipf's plane}, defined by the axes $Ns_m^{1/\gamma}$ and $s_M^{1/\gamma}$. Fig.~\ref{fig:zipf_plane} shows how the systems introduced above and other we will study in detail in the following moves in the Zipf's plane. The color gradients from blue to yellow correspond to increasing values of $\log Q$. A trajectory from the yellow to the blue region corresponds to a Zipfian dynamics, while going in the opposite direction indicates that Zipf's law can be followed only temporarily and so spuriously. 		

The implications of Zipfian dynamics can be better understood considering as temporal variable the sum of sizes at a given point of the evolution, 
\[
    n=\sum_{k=1}^N S(k) .
\]
For example, in regard of cities, $n$ coincides with the total urban population, for what concerns books it represents the total number of words used in the sampling, while in the case of earthquakes it is directly related to the total energy released by all the events considered. Using $n$ as a temporal variable, the condition for the onset of a Zipfian dynamics can be written as 
\[
    \frac{dQ}{dn}\leq0 \to \frac{dN}{dn}R^{1/\gamma}+\frac{N}{\gamma}R^{1/\gamma-1}\frac{dR}{dn}\leq0
\]
where we defined $R=s_m/s_M$. This yields 
\begin{equation}
    \frac{d\log\ton*{\frac{s_M}{s_m}}}{dn}\geq\gamma\frac{d\log N}{dn}
    \label{eq:coherence}
\end{equation}
This dynamical constraint, that we call \textit{coherence}, relates the growth of the probabilistic space (left side) to the growth of the physical space (right side), regarded as the number of groups or elements $N$ into which $n$ is partitioned. For systems showing Zipfian dynamics, a fast increase of $N$ must be compensated by an even faster growth of the range of the PDF.
This implies that Zipfian systems are out of equilibrium driven, because even if the system evolves and the number of elements enlarges, the underlying distribution is never completely sampled. In other words, for Zipfian systems, the empirical frequencies do not coincide with the generating PDF and this is a direct consequence of coherence. Indeed the growth of the probabilistic space is faster than the enlargement of the physical one, making the upper cutoff grow faster than the rate at which the probabilistic space is explored. Conversely, if the dynamics is not Zipfian and the system shows Zipf's law only temporally and so spuriously, the system evolves toward equilibrium: the probabilistic space is fully explored and the empirical frequencies become a good approximation of the inherent PDF. We can then divide the systems showing Zipf's law in two distinct categories, which show radically different dynamical properties:
\begin{itemize}
	\item \textbf{Spurious Zipfian systems} which present Zipf's law only at a certain moment of their evolution due to a temporary and accidental under-sampling. For these systems, Zipf's law does not represent an attractor of the dynamics and therefore this scaling law can not be used for characterizing their behavior. In this case the eventual observation of Zipf's law is entirely attributable to the underlying scale free distribution;
	\item \textbf{Genuine Zipfian systems} which dynamically evolve toward Zipf's law. Such behavior is produced by a fast enlargement of the probabilistic space, which makes these systems be out of equilibrium driven. In this case, the underlying scale free distribution is not sufficient to explain the stability of Zipf's law, indeed there must also be a dynamical constraint, that we call coherence, making the parameters of the PDF evolve in an appropriate way.
\end{itemize}
These different dynamical behaviors can be both visualized through the Zipf's plane, Fig.~\ref{fig:zipf_plane}, and by considering the evolution of $Q$ with respect to $n$, as shown in Fig.~\ref{fig:n_Q}. In this last case, red trajectories represent spurious Zipf's law, while blue ones are indicative of a Zipfian dynamics, so of genuine Zipf's law. The figure well demonstrates the need of approaching Zipf's law from a dynamical point of view, because in this way systems which could appear, at first glance, statistically similar, show all their differences.

\subsection{Relation with Heaps' law}\label{subsec:heaps}
\begin{figure*}
	\centering
	\includegraphics[width=1\linewidth]{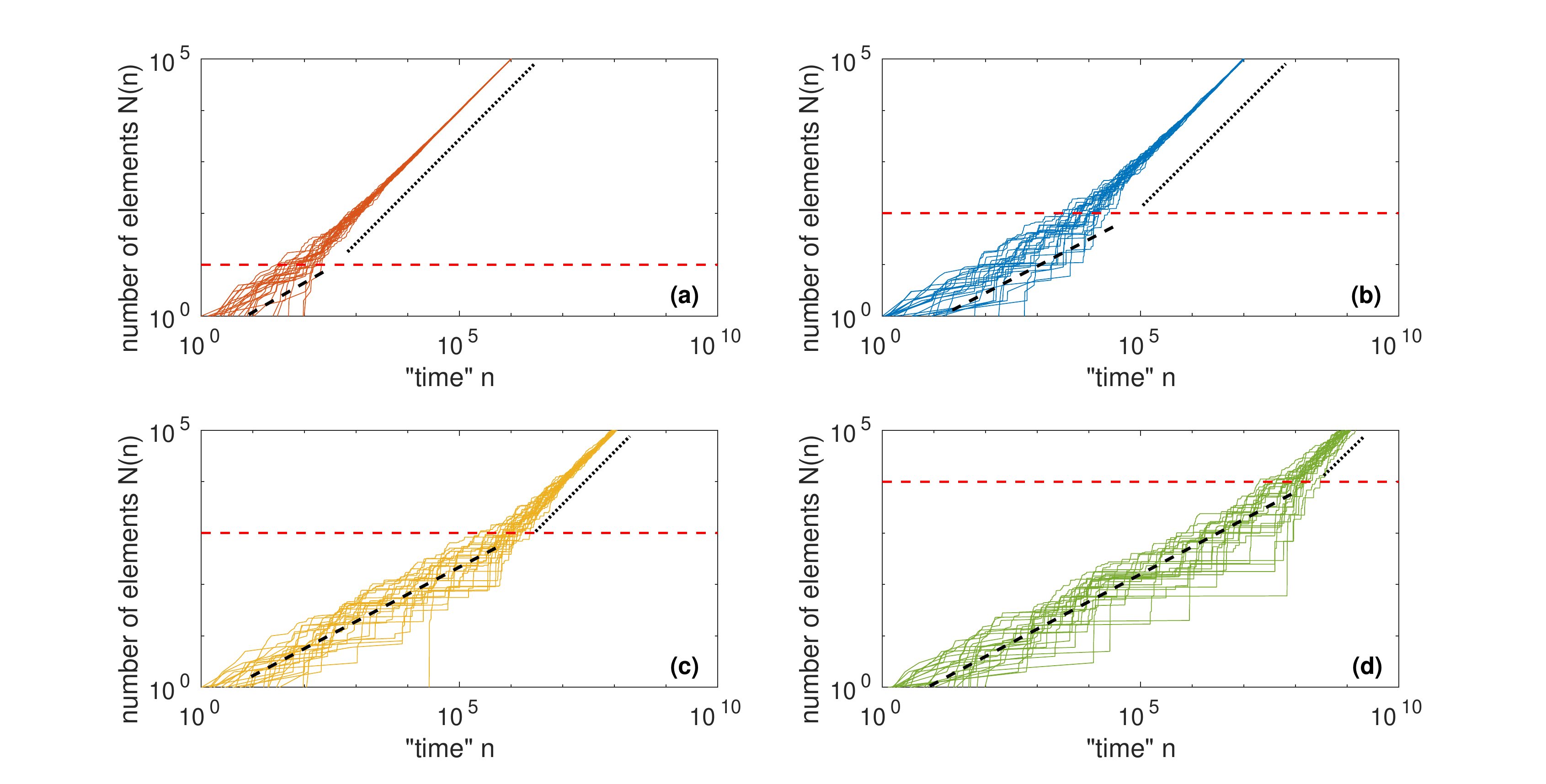}
	\caption{\textbf{Heaps' law and spurious Zipf's law.} Effects of a finite upper cutoff $s_M$ on the growth of $N(n)$. If the Zipf's exponent $\gamma$ is larger than one and the dynamics is not Zipfian, two scaling regimes are present. For $N\ll\ton*{\frac{s_M}{s_m}}^{1/\gamma}$, that is $Q\ll1$, it holds $N\sim n^{1/\gamma}$, while for $Q\gg1$ the growth of $N(n)$ is linear in $n$. Here we represented these different behaviors for $\gamma=2$, by performing a random sampling from a power law distribution with $\alpha=\frac{3}{2}$ and different upper cutoffs, namely $s_M=10^2$ (panel \textbf{(a)}), $s_M=10^4$ (panel \textbf{(b)}), $s_M=10^6$ (panel \textbf{(c)}) and $s_M=10^8$ (panel \textbf{(d)}). Red dashed lines represent the transition point between the two regimes, which are enlightened by the black dashed lines ($N(n)\sim n^{1/2}$) and by the black dotted ones ($N(n)\sim n$).}
	\label{fig:heaps_sM}
\end{figure*}
The dynamics of $N$ with respect to $n$ is usually described in terms of Heaps' law. Lu et al \cite{lu2010} derived an expression relating the growth of $n$ to $N$, more precisely
\begin{equation}
    n(N)=\frac{N^{\gamma}}{1-\gamma}\qua*{N^{1-\gamma}-1}.
    \label{eq:lu_N_n}
\end{equation}
This relation holds under the assumption of a stable Zipf's exponent $\gamma$ and lower cutoff equal to one. In the limit $N\to\infty$ the scaling of $N$ thus satisfies
\begin{equation}
    \begin{cases}
        N(n)\sim n \quad \text{for}\quad \gamma<1\\
        N(n)\sim n^{\frac{1}{\gamma}}\quad \text{for}\quad \gamma>1.
    \end{cases}
    \label{eq:lu_scaling}
\end{equation}
Proceeding as done by Lu et al, but taking into account the presence of an eventually non-zero $Q$, we can write $n$ summing over the $N$ sizes of the elements composing the system 
\[
    n=\sum_{k=1}^N\frac{\bar{s}}{(k+Q)^{\gamma}}\approx\bar{s}\int_1^N\frac{dk}{(k+Q)^{\gamma}},
\]
Recalling that $\bar{s}=N^{\gamma}s_m$ and using the expression for $Q$ we obtain 
\begin{align}
    n(N)=\frac{N^{\gamma} s_m}{1-\gamma}&\left\{\qua*{N+N\ton*{\frac{s_m}{s_M}}^{1/\gamma}}^{1-\gamma}-\right.\nonumber\\
    &-\left.\qua*{1+N\ton*{\frac{s_m}{s_M}}^{1/\gamma}}^{1-\gamma}\right\}
    \label{eq:scaling_heaps_Q}
\end{align}
This expression has to be compared with that derived by Lu et al Eq.~\eqref{eq:lu_N_n}.

First of all, let us consider a system for which the cutoffs $s_m$ and $s_M$ are fixed and let us suppose $s_M\gg s_m$, in this case Eq.~\eqref{eq:scaling_heaps_Q} predicts the presence of two different regimes. For $N\ll\ton*{\frac{s_M}{s_m}}^{1/\gamma}$, so for $Q\ll1$, Eq.~\eqref{eq:scaling_heaps_Q} can be approximated as 
\[
	 n(N)\approx\frac{N^{\gamma} s_m}{1-\gamma}\qua*{N^{1-\gamma}-1}\quad \text{for}\quad N\ll\ton*{\frac{s_M}{s_m}}^{1/\gamma},
\]
which, apart for the presence of $s_m$, coincides with the expression derived by Lu et al. However, when the sampling level enlarges and $Q$ increases, the situation changes. Indeed for $N\gg\ton*{\frac{s_M}{s_m}}^{1/\gamma}$ we can rewrite Eq.~\eqref{eq:scaling_heaps_Q} as 
\begin{align*}
	n(N)\approx\frac{Ns_m}{1-\gamma}&\left\{\qua*{1+\ton*{\frac{s_m}{s_M}}^{1/\gamma}}^{1-\gamma}-\right.\\
	&-\ton*{\frac{s_m}{s_M}}^{(1-\gamma)/\gamma}\Bigg\}\quad \text{for}\quad N\gg\ton*{\frac{s_M}{s_m}}^{1/\gamma}.
\end{align*}
The conclusion is that, apart for an initial transient, the growth of $N(n)$ is linear for any $\gamma$ if the cutoffs are fixed. The crossover $N_c$ between the two regimes satisfies 
\[
    N_c=\ton*{\frac{s_M}{s_m}}^{1/\gamma}.
\]
Clearly an analogous conclusion holds also if the cutoffs are not fixed but the dynamics is not Zipfian. Indeed in this case $Q=N\ton*{\frac{s_m}{s_M}}^{1/\gamma}$ is growing and therefore the scaling identified by Lu et al holds only transiently for $Q\ll1$. This behavior is reported in Fig.~\ref{fig:heaps_sM}, where we plotted different Heapsian trajectories. More precisely we performed four random samplings from a power law with exponent $\alpha=\frac{3}{2}$ (which corresponds to a Zipf's exponent $\gamma=2$) and we studied the scaling of $N(n)$ considering the following upper cutoffs: $10^2$, $10^4$, $10^6$ and $10^8$ (from panel \textbf{(a)} to panel \textbf{(d)}). The theoretical crossover is represented by red dashed lines, while the black lines enlighten the two scaling regimes, namely $N(n)\sim n^{1/2}$ and $N(n)\sim n$. 
			
We can now turn to genuine Zipfian systems, for which, being the dynamics Zipfian, $Q$ does not increase with $n$. As a consequence if $N$ is growing and for $n$ sufficiently large, it will hold $N(n)\gg Q(n)$ and therefore Eq.~\eqref{eq:scaling_heaps_Q} reduces to
\begin{equation}
    n(N)=\frac{N^{\gamma}s_m}{1-\gamma}\qua*{N^{1-\gamma}-1}.
	\label{eq:n_N_asymptotic}
\end{equation}
This expression is analogous to the one derived by Lu et al, but there is an explicit dependence on $s_m$. This implies that the scaling defined by Eq.~\eqref{eq:lu_scaling} asymptotically holds, but only if the lower cutoff is constant in time. As a consequence the conclusion that any system showing a stable Zipf's exponent always presents Heaps' law is not correct. Nevertheless Zipf's scaling is more fundamental than Heaps' one, as suggested by Lu et al.\cite{lu2010}, even if the relation among these laws is more complicated than previously noticed. In particular 
\begin{itemize}
	\item if the system shows spurious Zipf's law and the lower cutoff is fixed, then Heaps' exponent is asymptotically equal to one, independently of the Zipf's exponent. For $\gamma>1$ a transient regime is present, and it lasts up to $Q\sim1$. In this case, Heaps' exponent coincides with that found by Lu et al.\cite{lu2010};
	\item if the system shows Zipfian dynamics and so a genuine Zipf's scaling, then Heaps' law is found and Heaps' exponent asymptotically (i.e., as soon as $N\gg Q$) coincides with that derived in \cite{lu2010}, provided that the lower cutoff $s_m$ does not vary over time;
	\item if the lower cutoff $s_m$ is varying in time, the scaling of $N(n)$ can not be easily derived, being explicitly influenced by that of $s_m$. However, moving to the reference frame in which the lower cutoff is held fixed, this case can be traced back to those discussed above.
\end{itemize}
This explains, for instance, why the Sample Space Reducing process shows, for large $n$, a stable Zipf's law without the presence of Heaps' scaling \cite{mazzolini2018heaps}. Moreover it also clarifies the findings in \cite{lu2010} concerning the increasing of the Heaps' exponent in presence of an exponential cutoff in the tail of the generating power law. 

\subsection{Spurious Zipf's law and the upper cutoff}
We previously noticed that while a system showing Zipfian dynamics is out of equilibrium driven, never sampling the inherent PDF, spurious Zipfian systems completely explore the full range of the distribution during their evolution. As we are going to show, this property can be used to give an estimate of the upper cutoff $s_M$ for these systems. We recall that the deviation parameter satisfies
\[
    Q=N\ton*{\frac{s_m}{s_M}}^{1/\gamma},
\]
while, using Eq.\eqref{eq:rank_size}, we see that the largest object in the system is given by
\[
    S(1)=\frac{N^{\gamma}s_m}{(1+Q)^{\gamma}}.
\]
It then follows 
\[
    1+N\ton*{\frac{s_m}{s_M}}^{1/\gamma}=N\ton*{\frac{s_m}{s(1)}}^{1/\gamma}.
\]
This yields
\begin{equation}
	1+Q=Q_e
	\label{eq:Q_Q_e}
\end{equation}
where we defined the empirical deviation parameter $Q_e$, whose expression is 
\[
    Q_e=N\ton*{\frac{s_m}{s(1)}}^{1/\gamma}.
\]
We can then relate the upper cutoff of the PDF to the rank one object, in particular we obtain 
\[
    s_M=s_m\ton*{\frac{N}{Q_e-1}}^{\gamma}.
\]
For $Q_e=1$, that is for a perfect Zipf's law, the upper cutoff diverges, meaning that we can not infer it starting from the data. However, if the dynamics is not Zipfian, $Q$ increases with time and, as shown by Eq.~\eqref{eq:Q_Q_e}, $Q_e$ does the same. As a consequence, for $N$ sufficiently large, it will hold $Q_e\gg1$. In this limit we can expand the previous expression, obtaining
\begin{align}
    s_M&\approx s_m\qua*{\frac{N}{Q_e}\ton*{1+\frac{1}{Q_e}}}^{\gamma}=s(1)\ton*{1+\frac{1}{Q_e}}^{\gamma}=\nonumber\\
    &= s(1)\ton*{1+\frac{1}{1+Q}}^{\gamma}.
    \label{eq:s_M_S(1)}
\end{align}
Using this expression we can estimate the upper cutoff $s_M$ for any system showing spurious Zipf's law.

\section{Practical applications} 
In the following we apply the methodology introduced in the previous section to earthquakes, cities, and language. In all three cases our dynamical approach will provide novel and quantitative insights into the considered systems. Then we discuss two models which notoriously lead to Zipf's scaling: multiplicative processes \cite{levy1996power} and the Yule-Simon model \cite{yule1925ii, simon1955class}, showing analytically that they produce a truly Zipfian dynamics.

\subsection{Earthquakes}
\begin{figure*}[t]
	\centering
	\includegraphics[width=\linewidth]{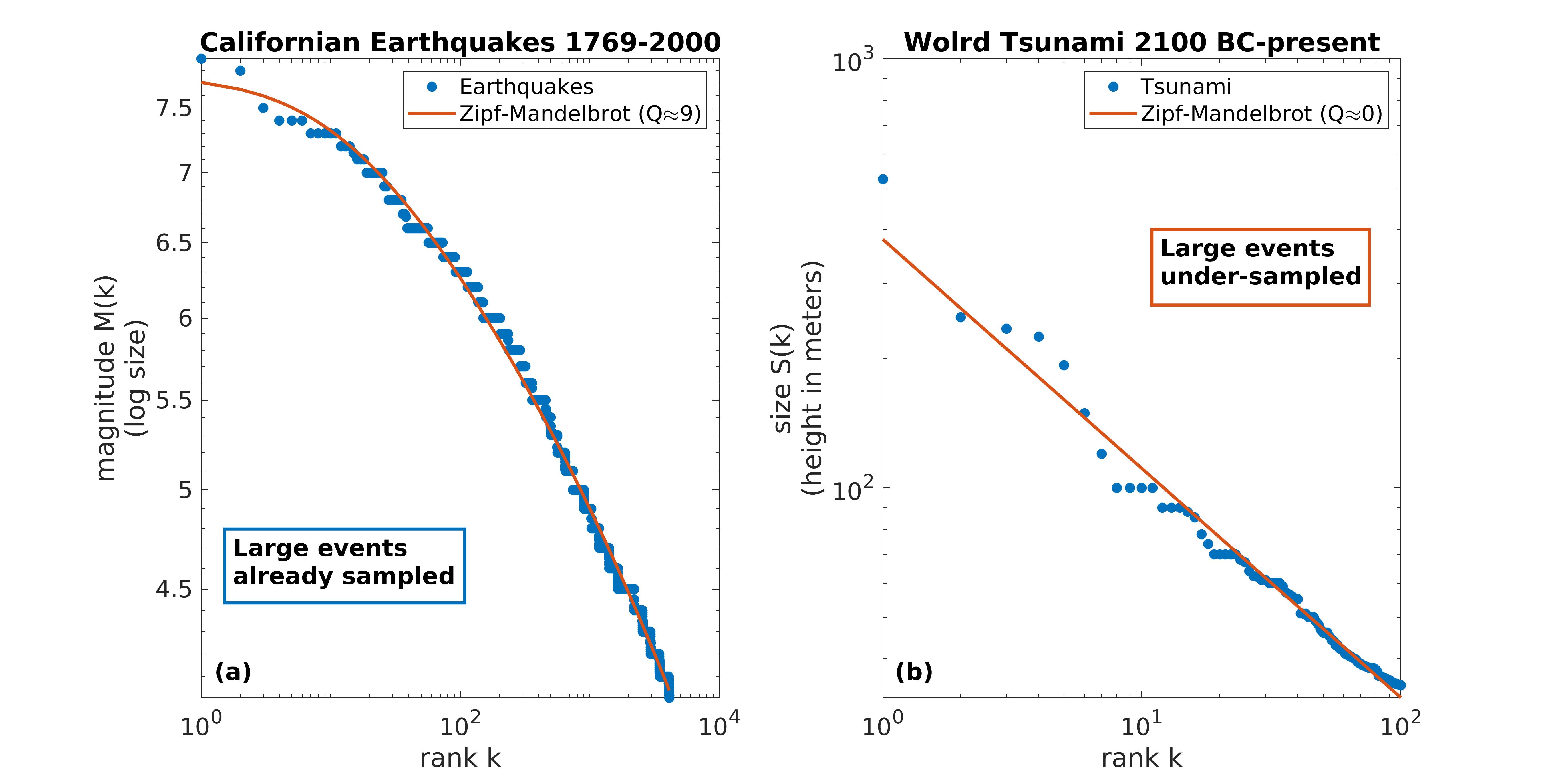}
	\caption{\textbf{Natural hazards.} \textbf{(a)} Rank-size plot of Californian earthquakes registered between 1769 and 2000 and with magnitude larger than 4. The larger deviations with respect of the smaller time interval analyzed in \cite{Newman} indicate the lack of Zipfian dynamics. Moreover such deviations imply that the strongest earthquake observed is a good approximation of the upper cutoff. \textbf{(b)} Rank-size plot of world tsunami occurred from 2100 BC to 2020. In this case considering a large time window does not make deviations appear. The conclusion is that the available sample does not allow to infer the upper cutoff of tsunami height, which, as in the case of earthquakes, is expected to vary only over geological scales.}
	\label{fig:earthquakes_tsunami_Q}
\end{figure*}

It is well known that earthquakes follow the Gutenberg-Richter law \cite{Gutenberg} - i.e. the energy released is power-law distributed - and it is reasonable to assume that, in a given seismic zone, the upper cutoff of this PDF can only vary over geological times. By using Eq.~\eqref{eq:bar_s_Q}, we thus deduce that increasing the numerosity $N$ of the set, that is by considering larger and larger time windows (but always much smaller than geological scales), results in higher values of $Q$. This is confirmed by the corresponding trajectory in the Zipf's plane, Fig.~\ref{fig:zipf_plane}, where we plotted the trajectory of Italian earthquakes. From right to left, the points correspond to an interval of $50$ years ($1950-2000$), of $100$ years ($1900-2000$) and so on, up to $1000$ years ($1000-2000$). These points accumulate in correspondence of the maximum possible size for an earthquake occurring in Italy, enlightening the absence of a Zipfian dynamics. This is also confirmed by the growth of $Q(n)$, represented in Fig.~\ref{fig:n_Q}. The conclusion is that earthquakes can show only spurious Zipf's law. This result directly derives from the fact that earthquakes, neglecting short time correlations, are by a good extent independent over long periods (tens of years), being energy always injected into the system. It is therefore clear that power-law distributed objects cannot tend to Zipf's law if they evolve independently, as long as the cutoffs of the inherent PDF are fixed. Moreover, by looking at \ref{fig:earthquakes_MobyDick_Cities}, it is possible to conclude that no future Italian earthquake will be substantially stronger than the largest event already recorded, which is a rather interesting and non-trivial result, being obtained only from simple statistical considerations. Indeed, the deviation parameter $Q$ is large and we can use Eq.~\eqref{eq:s_M_S(1)} for obtaining an estimate of the maximal magnitude of an earthquake occurring in Italy
\[
    M_{max}^{it}\approx 7.4
\]
			
These considerations also explain the findings of Newman \cite{Newman} and Sornette et al \cite{sornette1996rank} about Californian earthquakes. Newman analyzes only events recorded in the period 1910-1992, finding a pure Zipf's law with no deviations. Analogously Sornette et al considered earthquakes occurred in Southern California in the period 1930-1990. In both cases no deviations from Zipf's law are observed and this is due to the small dimension of the samples used. Indeed, the time window considered is too small to appreciate the upper cutoff and deviations appears increasing the sampling interval. This is shown in Fig.~\ref{fig:earthquakes_tsunami_Q}, were we plotted the rank-size plot of Californian earthquakes occurred between 1769 and 2000 with the corresponding fit. Being $Q\approx9$ we conclude that also in this case the largest earthquake registered in the sample we considered, whose magnitude is $M(1)^{ca}\approx7.9$, is a good estimator of the upper cutoff of the earthquakes size distribution.
			
Our dynamical approach consequently shows that a critical usage of the rank-size plot provides information which goes well beyond the simple identification of a scale free distribution. Indeed the presence of deviations at low ranks and the identification of spurious Zipf's law allow to perform risk assessment and to understand if the available data are a sufficient statistic of the phenomenon considered. For instance, if the rank-size plot of earthquakes occurred in a given region is straight, then the most powerful earthquake observed will not be, in general, a good estimate of the upper cutoff of the distribution. Clearly this has strong implications for the planning of antiseismic measures. As a consequence, when studying the risk connected with natural hazards, an inspection of the rank-size plot can be a very fast procedure to understand the effective reliability of the available data. For example, in respect of world tsunami, considering a very large time window does not make deviations from Zipf's law appear. This is shown in Fig.~\ref{fig:earthquakes_tsunami_Q}, where we plotted the rank-size plot of tsunami occurred worldwide since $2100$BC. For tsunami run-up heights we used NOAA NCEI/WDS Global Historical Tsunami Database, 2100 BC to Present \cite{noaaTsunami}. Also in this case the upper cutoff of the distribution is expected to be fixed over human times, therefore the conclusion is that there is no statistical evidence that the highest tsunami ever observed represents a good estimate for the upper cutoff of tsunami distribution.

\subsection{Cities}
\begin{figure*}[t]
    \centering
	\includegraphics[width=\linewidth]{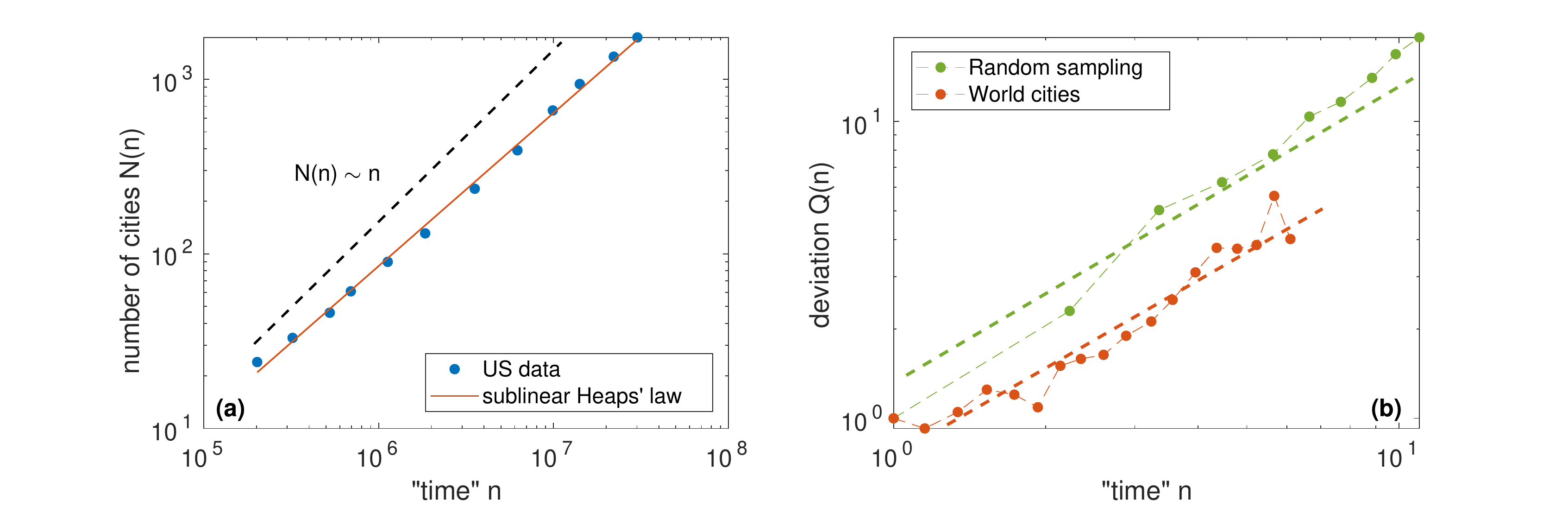}
    \caption{\textbf{(a)} Growth of the number of US cities as function of the urban population. Being the dynamics of US cities Zipfian and the size of the smallest urban settlement constant we expect the system to show Heaps' law. Note that this growth is sub-linear as expected from the finite size of the system considered. \textbf{(b)} Evolution of $Q(n)$ as function of "time" $n$ for a random sampling and world cities. Dashed lines represent the trend $Q(n)\sim n$. As predicted by Eq.~\eqref{eq:Q_n_growth} $Q(n)$ grows linearly in $n$ for the random sampling. Remarkably this linear growth is observed also considering world cities. Being the lower cutoff of such system fixed (and equal to $300,000$ inhabitants), this suggest that the largest urban centers are reaching the intrinsic upper cutoff of the distribution.} 
    \label{fig:cities_all}
\end{figure*}
\subsubsection{US cities are Zipfian}

The population of metropolitan areas is a prototype of system usually claimed to follow Zipf's law. Here we show that, while US cities follow a Zipfian dynamics, when we consider world cities the dynamics is not Zipfian, a direct consequence of the non-additivity property of Zipfian dynamics, that we analytically prove below.\\
Let us first focus on the time evolution of US cities. Up to 1776 the US were, by a good extent, independent entities and each of them used its resources to make its own capital grow. As soon as interaction became relevant and the USA turned into a single nation, resources were centralized, flowing into only some of those cities and allowing them to reach a population that would have been impossible to sustain for a single State. As a consequence, the upper cutoff of the size distribution of cities enormously increased. This process, which corresponds to the emergence of New York as the driving city of the USA, is well represented by the corresponding trajectory in the Zipf's plane (Fig.~\ref{fig:zipf_plane}, black circles). The size of the largest possible city $s_M^{1/\gamma}$ increases very fast with respect to $N\cdot s_m^{1/\gamma}$, leading to a decrease of $Q(n)$, as shown in Fig.~\ref{fig:n_Q}. \footnote{Note that the size of the smallest urban settlement, $s_m$, is by a good extent constant: even if existing cities grow, new ones are constantly founded, avoiding the growth of the lower cutoff.} \\ 
Being the dynamics Zipfian, we can that expect US cities to follow also Heaps' law. Indeed the lower cutoff of the distribution, which coincides with the size of the smallest urban settlement, remained almost constant during the development of US urban system, being administratively fixed at $2500$ inhabitants. The applicability of Heaps' law to urban structures has been widely ignored and only recently this point has been considered \cite{simini}, even if not from a dynamical point of view. More precisely previous works \cite{simini} study the functional form of $N(n)$ by plotting the number of cities as function of the population for many countries at present days. This yield a static and aggregate picture, which lacks in providing any information about the urban development of a given nation. Differently, here we focus on the dynamics of a single country, namely the US, following its urban development over time. We reported in panel \textbf{(a)} of Fig.~\ref{fig:cities_all} the growth of the number of US cities $N(n)$ as function of the urban population $n$, also a fit to Heaps law is drawn. The adherence to this scaling relation is strict and the exponent $\beta$ of Heaps' scaling satisfies
\[
    \beta=0.8768 \ (0.8459, \ 0.9077)
\]
This sub-linear growth is what one would expect being the size of the system finite and the average (over the period 1790-1900) Zipf's exponent $\gamma\approx0.86$ \cite{lu2010}. The fact that also urban systems evolve according to Heaps' law has never been pointed out, but it is a very natural consequence of the dynamical framework we developed.

\subsubsection{World cities are not Zipfian}

Now we consider an aspect of Zipf's law that, despite its relevance, has been only partially discussed \cite{Batty}: is the union of two Zipfian set still Zipfian? In this section we prove that Zipfian dynamics is not an additive property. Urban systems constitute a perfect framework to apply our framework to fully understand this phenomenon. Zipf's law is observed for almost any country \cite{soo2005zipf} and therefore, also guided by our findings regarding the US, one could expect that also the system formed by all cities in the world is Zipfian. For each individual country $k$ coherence is respected, consequently we have
\[
    \frac{d\log\ton*{\frac{s_M^{(k)}}{s_m^{(k)}}}}{dn^{(k)}}\geq\gamma^{(k)}\frac{d\log N^{(k)}}{dn^{(k)}}
\]
where the apex $(k)$ indicates that all these quantities are referred to the $k$th nation. Denoting by $M$ the number of countries, the system formed by world cities is obtained summing over the $M$ national urban system, this implies 
\[\begin{cases}
    N=\sum_{k=1}^MN^{(k)}\\
    n=\sum_{k=1}^Mn^{(k)}
\end{cases}\]
where $N$, the physical space, is the number of cities in the world and $n$ the total urban population. The lower cutoff $s_m^{(k)}$ is clearly country independent and so we can set $s_m^{(k)}=s_m$ for any $k$. Differently, the upper cutoff of the world system $s_M^{(k)}$ coincides with the largest $s_M^{(k)}$, let us say $s_M^{(a)}$. Note that this cutoff will not be, in general, the one growing faster. We finally assume that the Zipf's exponent is approximately the same for all countries i.e. $\gamma^{(k)}\approx\gamma$ for any $k$. This approximation is well supported by empirical studies \cite{soo2005zipf}. We can now write the coherence condition for the world system: the right side of Eq.~\eqref{eq:coherence} is
\[
    \gamma\frac{d\log N}{dn}=\frac{\gamma}{N}\sum_{k=1}^M\frac{dN}{dn^{(k)}}\frac{dn^{(k)}}{dn}=\frac{\gamma}{N}\sum_{k=1}^M\frac{dN^{(k)}}{dn^{(k)}}
\]
Dividing and multiplying by $N^{(k)}$ and introducing the frequencies $x^{(k)}=\frac{N^{(k)}}{N}$ we rewrite this expression as 
\[
    \gamma\sum_{k=1}^Mx^{(k)}\frac{d\log N^{(k)}}{dn^{(k)}}=\gamma\mean*{\frac{d\log N^{(k)}}{dn^{(k)}}}_k
\]
where $\mean{\cdot}_k$ is the average over the different countries. The left side of Eq.~\eqref{eq:coherence} is instead 
\[
    \frac{d\log\ton*{\frac{s_M}{s_m}}}{dn}=\frac{d\log\ton*{\frac{s_M^{(a)}}{s_m}}}{dn}=\frac{d\log\ton*{\frac{s_M^{(a)}}{s_m}}}{dn^{(a)}}
\]
We then obtain that coherence is found if it holds
\[
    \frac{d\log\ton*{\frac{s_M^{(a)}}{s_m}}}{dn^{(a)}}\geq\gamma\mean*{\frac{d\log N^{(k)}}{dn^{(k)}}}_k
\]
This implies that the system formed by world cities is Zipfian only if the growth of the probabilistic space characteristic of the country with the largest upper cutoff is bigger than the average growth of the physical space. Clearly this condition is not a direct consequence of the coherence of country $(a)$ and, indeed, world countries do not show a Zipfian dynamics, as confirmed by the corresponding trajectory in the Zipf plane Fig.~\ref{fig:zipf_plane} and the evolution of $Q(n)$ in Fig.~\ref{fig:n_Q}. The historical population of world cities comes from the World Urbanization Prospects 2018 of the Department of Economic and Social Affairs (UN) \cite{world_cities}. In particular we used the "Annual Population of Urban Agglomerations with $300,000$ Inhabitants or More in 2018, by country, 1950-2035" (ignoring 2021-2035 data).

The dynamics of $Q$ for world cities is shown in more detail in panel \textbf{(b)} of Fig.~\ref{fig:cities_all}, where we plotted also the trend which results from a random sampling with fixed cutoffs. Note that $Q$ and $n$ have been rescaled for better comparing the two systems and moreover we used only values $Q\gtrsim1$ for avoiding the effect of noise on small $Q$. In the case of the random sampling, $Q(n)$ is expected to be a linear function of $n$, indeed, recalling Eqs.~\eqref{eq:bar_s_Q} and the results of Subsec.~\ref{subsec:heaps} regarding spurious Zipfian systems, it holds
\begin{equation}
    Q(n)=N\ton*{\frac{s_m}{s_M}}^{1/\gamma}\sim n\ton*{\frac{s_m}{s_M}}^{1/\gamma}.
    \label{eq:Q_n_growth}
\end{equation}
Being the cutoffs and the Zipf's exponent fixed, $Q$ grows linearly with $n$, as also confirmed by panel \textbf{(b)} of Fig.~\ref{fig:cities_all}. Remarkably this trend is observed also considering world cities, as shown in the same figure. Being $s_m$ fixed (and equal to $300,000$), this suggest that also in this case $s_M$ is fixed, meaning that the largest cities in the world are getting closer and closer to an upper limit of population. This is consistent with many studies asserting that large urban centers are not efficient due to, for instance, traffic jams, pollution, complexity of water management or vulnerability to natural hazards \cite{molina2004megacities, varis2006megacities, wenzel2007megacities}. Exploiting Eq.~\eqref{eq:s_M_S(1)} we can then give an estimate of the upper cutoff of urban population, it results 
\[
    s_M\approx 41\cdot10^6\ \text{inhabitants}.
\]
This number has to be compared with the population of Tokyo metro area, that, with a population of $\approx 37\cdot 10^6$ inhabitants, is the largest urban settlement in the world. Clearly, in contrast with earthquakes, the upper cutoff of world cities may vary thanks to, for instance, technological innovations, as happened with the development of skyscrapers in the past century. As a consequence, the value we computed should be considered as a limit of population which is expected not to be overcome in the next few years. In this sense our estimate, obtained only by statistical arguments, is in good agreement with projections \cite{hoornweg2017population}, according to which this population limit will substantially hold up to $2050$.
\subsection{Language}

\subsubsection{The dynamics of language}      	
Natural language is the first and most prominent application of Zipf's law. We can confirm that this system is Zipfian by looking the evolution of $Q(n)$ Fig.~\ref{fig:n_Q} and the trajectory in the Zipf's plane Fig.~\ref{fig:zipf_plane}, both referred to the words occurrences in the Moby Dick novel. Here the objects are the different words, the sizes $S$ are their numbers of occurrence, and ``time" $n$ is the progressively increasing fraction of words of the novel we consider. In this case, occurrences of different words are not independent events, being them constrained by grammar and semantic rules, which makes the upper limit of the number of occurrences grow faster than the product $N^{\gamma}\cdot s_m$. We recall that $N$ is the number of different words observed. For instance, in order to increase $N$, new meaningful sentences, semantically coordinated with the previous text, are to be composed. However these sentences must contain, on average, many occurrences of the most frequent word, which is the article "the". This makes $s_M$ grow faster than $N^{\gamma}\cdot s_m$ and thus a coherent, Zipfian dynamics emerge. 
			
In order to prove that Zipfian dynamics emerges as a consequence of the effect of correlations induced by grammar and semantic rules, we consider two different systems in which such rules are adopted to a different extent. In panel \textbf{(a)} of Fig.~\ref{fig:language_all} we show the rank-size plot of the most common words used by children and adults. In particular we used CHILDES database \cite{macwhinney2000} and we analyzed with CLAN program all the American English corpora available to obtain two rank-size lists, one referred to children below six and the other to adults. It is evident that childish language is characterized by considerable deviations from Zipf's law and so by a larger value of $Q$, in particular performing two fits we obtained $Q_{children}\approx3.6$ and $Q_{adults}\approx0.80$. Language, seen as a system which evolves during its learning, is then characterized by a Zipfian dynamics.  

\subsubsection{Zipfian dynamics increases language efficiency}  \begin{figure*}[t]
	\centering
	\includegraphics[width=\linewidth]{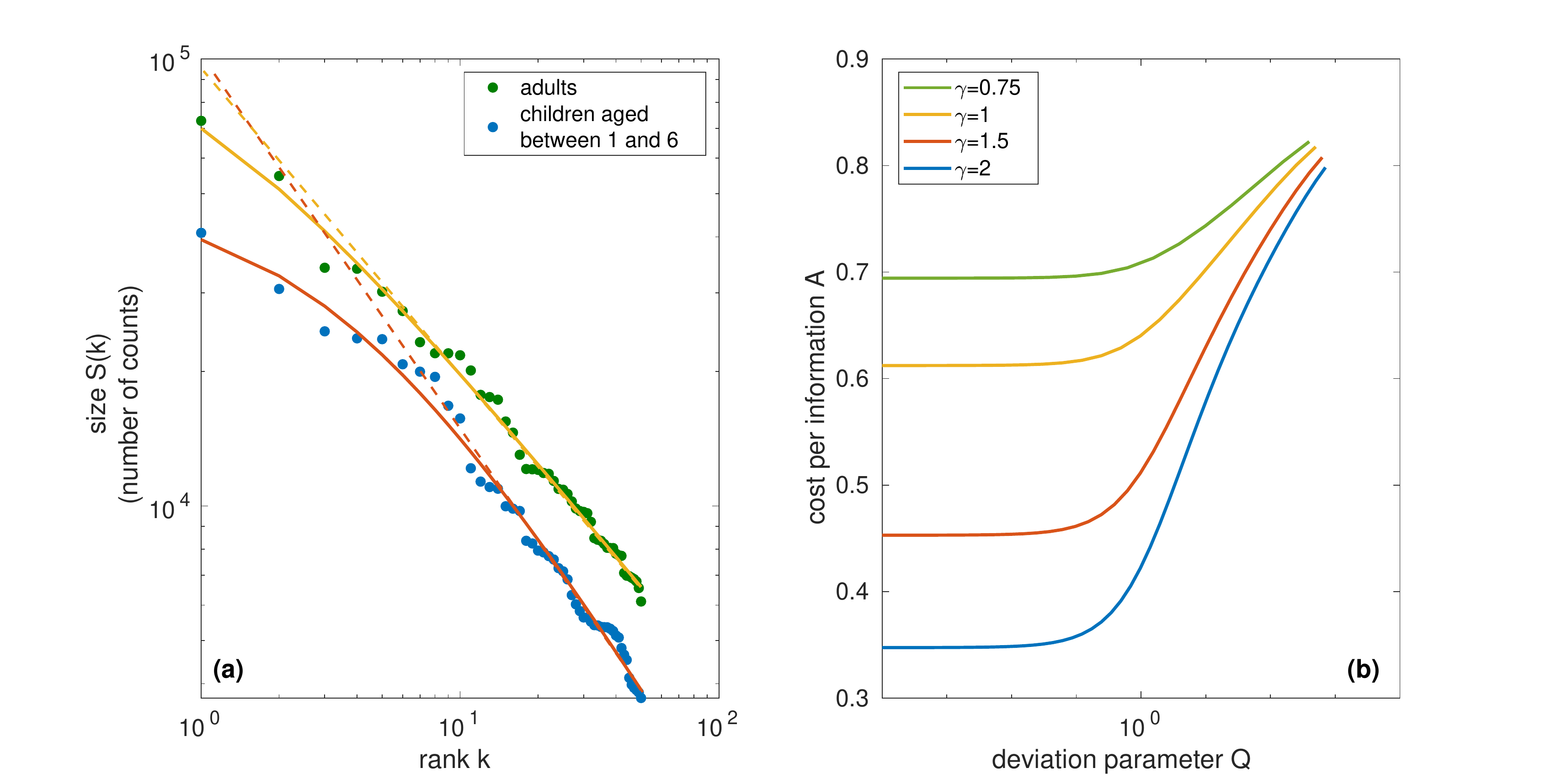}
	\caption{\textbf{Dynamics of language.} \textbf{(a)} Rank-size plot of the most common $50$ words used by children and adults. Deviations from Zipf's law are much more prominent in the former, a consequence of the imperfect adoption of grammar rules at the early age. This suggests that the dynamics of language is Zipfian. \textbf{(b)} Average cost per information of simulated languages as function of the deviation parameter $Q$. As it is possible to see, at fixed Zipf's exponent $\gamma$, lower is $Q$, larger is the efficiency of the language (inverse of $A$). The Zipfian dynamics of language can then be explained in terms of an optimization process.}
	\label{fig:language_all}
\end{figure*}	   	

A well know argument explaining the onset of Zipf's law in language is due to Mandelbrot \cite{Mandelbrot, mitzenmacher2004brief} and it is based on information theory. Let us consider a dictionary of $N$ words, each of them will be characterized by a frequency of occurrence which can be regarded as its normalized size. If we want to efficiently store and transmit sentences the best thing to do is to associate low coding numbers to the most common words. Denoting by $C(k)$ the coding of the $k$th word by frequency the most efficient choice is 
\[
    C(k)=\log_2(k)
\]
The average information per word is given by the entropy of the language $H$, defined as 
\[
    H=-\sum_{k=1}^{N}f(k)\log_2f(k)
\]
where $f(k)$ is the frequency of the $k$th most common word. As pointed out by Mandelbrot, an efficient language should maximize the average information, while lowering the average cost $C$, whose expression is 
\[
    C=\sum_{k=1}^NC(k)f(k)
\]
In other words, the quantity $A=C/H$ must be as small as possible. By minimizing $A$ one obtains that $f(k)$ follows Zipf's law \cite{mitzenmacher2004brief, Mandelbrot}, but the Zipf's exponent $\gamma$ depends on the vocabulary size $N$ and diverges for large $N$ \cite{manin2009mandelbrot}. This drawback seriously compromise the argument as originally formulated, however Mandelbrot's idea can still be used for understanding Zipfian dynamics in language. Indeed, a large $Q$ implies that the empirical distribution $f(k)$ is curved at low ranks, making the most common words being equally frequent. This is expected to lower the efficiency of the language, as follows from the expression of the cost $C$. We have numerically proved this by studying the value of $A$ as function of $Q$ at fixed Zipf's exponent. In particular, we extracted $N$ frequencies using different values of the upper cutoff $s_M$ and we repeated the process for various Zipf's coefficients $\gamma$. Results are reported in panel \textbf{(b)} of Fig.~\ref{fig:language_all}. It clearly emerges that, ceteris paribus, the lower is $Q$ the lower is $A$. As a consequence the presence of a Zipfian dynamics in natural language has a very natural explanation in the context of linguistic and information theory, because an evolution towards low values of $Q$ increases the efficiency of the language in terms of the ratio between information and cost. In other words, even if Mandelbrot's argument does not explain why a finite Zipf's exponent is observed, it allows to understand why, if Zipf's exponent is hold fixed by some other mechanism, the language evolves with a Zipfian dynamics. 

\subsection{Multiplicative process}
Multiplicative processes are a well known example of models capable of explaining the onset of power laws \cite{levy1996power}; in particular, they have been used to model the evolution of incomes and stock prices \cite{Multiplicative, mitzenmacher2004brief}. Given the variable $S_t$, representing for instance the price of a stock at time $t$, its evolution is defined as 
\[
    S_{t+1}=r_t\cdot S_t
\]
where $r_t$ is a random variable drawn at $t$ from a given PDF, usually a normal or a uniform distribution. This stochastic dynamics is equivalent to a random walk in logarithmic space and it is easy to show that, under mild assumptions, the variable $S$ is asymptotically log-normally distributed \cite{redner1990random}. If however we constrain $S_t$ to be larger than a lower cutoff $s_m$ (that is, $S_t$ performs a random walk with a reflecting barrier in $s_m$), then the limiting distribution of $S$ becomes a power law \cite{levy1996power}. Making the continuum approximation the process can be described in terms of a Fokker-Planck equation with drift $\nu$ and diffusion coefficient $D$. In particular it holds \cite{sornette1997convergent}
\[
    \begin{cases}
        \nu=\mean*{\log r}\\
        D=\mean*{\ton*{\log r}^2}-\mean*{\log r}^2
    \end{cases}
\]
and the limiting distribution of $S$ is a power law of the form 
\[
    P(S)\sim S^{-\mu-1}\ \text{with}\ \mu=\frac{|\nu|}{D}.
\]
In the transient regime, however, the power law distribution above presents an upper cutoff $s_M$ given by \cite{sornette1997convergent}
\begin{equation}
    s_M(t)=\me^{\sqrt{Dt}} .
    \label{eq:s_M_multiplicative}
\end{equation}
In other words, the probabilistic space enlarges exponentially fast with time. Now we show that this leads to a Zipfian dynamics.

Let us consider an ensemble of $N$ objects, which could be, for instance, a set of stock prices, evolving according to the multiplicative process with lower cutoff described above. Suppose that all the prices are initially equal to $s_m$, Recalling the expression of $Q$ Eq.~\eqref{eq:bar_s_Q} and using Eq.~\eqref{eq:s_M_multiplicative}, we obtain that the deviation parameter $Q$ evolves in time according to
\[
    Q(t)=Ns_m^{1/\gamma}\me^{-\sqrt{Dt}/\gamma}.
\]
So $Q$ exponentially decreases towards zero and the system shows Zipfian dynamics. The coherence condition Eq.~\eqref{eq:coherence} can be written as
\[
    \frac{d\log\ton*{\frac{s_M}{s_m}}}{dn}=\frac{d\log\ton*{\frac{s_M}{s_m}}}{dt}\frac{1}{\frac{dn}{dt}}\geq\gamma\frac{d\log N}{dn}.
\]
Using the example of stock prices, here $n$ would be the total value of the stocks considered, being the sum of the $N$ prices. $N$ is fixed, so we rewrite this condition as 
\[
    \frac{d\log\ton*{\frac{s_M}{s_m}}}{dt}\frac{1}{\frac{dn}{dt}}\geq0
\]
The first derivative, by virtue of Eq.~\eqref{eq:s_M_multiplicative}, is positive, and the second is non negative, because all the prices are initially set equal to the lower cutoff $s_m$. We conclude that the multiplicative process is an example of Zipfian dynamics, as also shown by a typical trajectory in the Zipf plane Fig.~\ref{fig:zipf_plane} and the decrease of $Q(n)$ in numerical simulations Fig.~\ref{fig:n_Q}. Also, we note that even if the dynamics is Zipfian, the systems can not show Heaps' law because $N$ is constant. This could look like a contradiction, because we stated that Heaps' law asymptotically holds whenever the dynamics is Zipfian and the lower cutoff is fixed. A remark clarifies this point, after an initial transient, $n$ fluctuates around a constant value, as shown in Fig.~\ref{fig:n_Q}, because the ensemble reaches a stationary Zipfian distribution, whose parameters can be derived using Eq~\eqref{eq:bar_s_Q}
\[
	S(k)=\frac{N^{\gamma}s_m}{k^{\gamma}}
\]
As a consequence, being $N$, $s_m$ and $\gamma$ fixed, there is no growth of $n$ \footnote{We recall that $n=\sum_{k=1}^NS(k)$.} and then Heaps' law can not be defined at all. 

\subsection{Yule-Simon model}

Yule-Simon process \cite{yule1925ii, simon1955class}, based on the concept of preferential attachment, is one of the most famous examples of power laws generating model. We shortly illustrate the process in the contest of urban systems. Consider a set of $N_0$ initial urban centers, each with unitary population. At each time $n$:
\begin{itemize}
    \item with probability $1-p$ a unit of population is added to a random urban center $j$, selected with a probability proportional to its population $S_n(j)$;
    \item with probability $p$ a new urban settlement, with unitary population, is added to the system.
\end{itemize}
It has been proven that these cities are asymptotically power law distributed, more precisely 
\begin{equation}
    P(S)\sim S^{-\alpha} \quad \text{with} \quad \alpha=1+\frac{1}{1-p}
    \label{eq:Yule_Simon}
\end{equation}
Now we prove that the Yule-Simon model shows coherence. We start by noting that Eq.\eqref{eq:Q_Q_e} implies that if $Q_e$ is decreasing also $Q$ decreases. This consideration implies that coherence is found if it holds 
\begin{equation}
    \frac{d\log\ton*{\frac{S_n(1)}{s_m}}}{dn}\geq\gamma\frac{d\log N}{dn}.
    \label{eq:coherence_S(1)}
\end{equation}
In the context of urban settlements, $S_n(1)$ is the population of the largest city, $N$ is the number of different urban settlements, and $n$ is the total population. Moreover, the lower cutoff $s_m$ is equal to one (cities with unitary population are injected into the system at a constant rate). The population of the largest city, $S_n(1)$, evolves according to 
\[
    S_{n+1}(1)=S_n+(1-p)\frac{S_n(1)}{n} \to \frac{dS_n(1)}{dn}=(1-p)\frac{S_n(1)}{n},
\]
while the growth of $N$ satisfies 
\[
    N(n+1)=N(n)+p\to \frac{dN}{dn}=p, \quad N(n)=N_0+pn .
\]
Plugging these expressions into Eq.~\eqref{eq:coherence_S(1)}, and recalling Eq.~\eqref{eq:Yule_Simon}, we get 
\[
    \frac{1-p}{n}\geq(1-p)\frac{p}{N_0+pn}\to \frac{1}{n}\geq\frac{1}{n+\frac{N_0}{p}}
\]
That is always satisfied, since $N_0>0$ and $p>0$. We have thus proved that also Yule-Simon model satisfies the coherence condition and, therefore, that it shows Zipfian dynamics. This is also visually confirmed by the corresponding trajectory in the Zipf's plane Fig.~\ref{fig:zipf_plane}, which corresponds to a Yule-Simon process with $N_0=100$ and $p=0.5$, and by the evolution of $Q$ reported in Fig.~\ref{fig:n_Q}. Finally, it is interesting to note that such trajectory is very similar to the one performed by world countries, where the size of a country is given by its GDPppp. The rich get richer mechanism seems therefore more appropriate than a multiplicative process for describing the evolution of the world system.

\section{Conclusions and discussion}
\renewcommand{\arraystretch}{2.7}
\begin{table*}[]
    \centering
    \begin{tabular}{||c|c|c|c|c|c||}
    \hline
    \textbf{System} & \textbf{Size} $S$ & \textbf{Physical Space} $N$ & \textbf{Temporal variable} $n$ & \textbf{Zipfian Dynamics?} & \textbf{Main findings}\\
    \hline\hline 
    Earthquakes & \pbox{50cm}{Maximal amplitude\\(exponential of\\ the magnitude)} & number of earthquakes & total energy released & no & \pbox{20cm}{estimation of the \\ maximal magnitude}\\
    \hline
    US cities & population & \pbox{20cm}{number of urban\\ settlements} & total urban population & yes & \pbox{50cm}{US cities evolved \\ following Heaps' law}\\
    \hline
    World cities & population & \pbox{20cm}{number of urban\\ settlements} & world urban population & no & \pbox{20cm}{Zipfian dynamics \\ is not additive}\\
    \hline
    Language & number of counts & \pbox{30cm}{number of distinct\\ words} & total word used & yes & \pbox{20cm}{Zipfian dynamics \\ optimizes the language}\\
    \hline
    World countries & GDPppp & \pbox{20cm}{number of \\ countries} & world wealth & yes &  \pbox{20cm}{presence of a rich get\\ richer mechanism} \\
    \hline
    \pbox{20cm}{Multiplicative\\ process} & price & \pbox{20cm}{number of stocks} & \pbox{20cm}{total value of\\ the stocks} & yes & \pbox{20cm}{}\\
    \hline
    \pbox{20cm}{Yule-Simon\\ model} & population & \pbox{20cm}{number of urban\\ settlements} & total population & yes & \pbox{20cm}{Explain the trajectory\\ of world countries \\ }\\
    \hline
    \end{tabular}
    \caption{\textbf{Variables overview and main findings} Summary of the systems and models considered with the corresponding variables and the main findings obtained thanks to the dynamical approach.}
    \label{tab:my_label}
\end{table*}
Zipf's law is a scaling relation present in the rank-size plots of many different natural and socio-economic systems. Despite its ubiquity and the numerous empirical and theoretical investigations, a deep understanding and a unified framework of analysis is still lacking. In this work, we start from the empirical observation that the deviation parameter $Q$ of a given system changes with time, or even considering different subsets of the same database. The importance of this parameter is enhanced by its connection to the first ranks, i.e. the largest objects, the larger is $Q$, the more the first ranks deviate from Zipf's law. A probabilistic argument permits us to express the deviation $Q$ as a function of the intrinsic natural cut-offs of the underlying power law PDF ($s_m$ and $s_M$), its exponent $\alpha$, and the number of elements in the system $N$, Eq.~\eqref{eq:bar_s_Q}:
\[
    Q=N\ton*{\frac{s_m}{s_M}}^{1/\gamma}
\]
This relation permits to understand why a \textit{dynamical} approach is crucial: in general, for real systems $N$, $s_m$, and $s_M$ vary during evolution and, as a consequence, deviations from Zipf's law can increase, as in the case of earthquakes, or decrease, as happens with US cities. For this reason, we introduce the concept of \textit{Zipfian dynamics}, which drives the system to exact Zipf's law, and \textit{Zipf's plane}, a visual tool which allows studying deviations dynamically. In particular, we demonstrate that Zipfian dynamics can not be produced by a simple truncated power law PDF of sizes. Indeed, it is connected to the presence of mechanisms which make the level of sampling and the parameters of the PDF evolve in a peculiar way, such that $Q$ decreases during evolution. More precisely, we find a dynamical constraint, that we name \textit{coherence}, relating the growth of the probabilistic space $s_M/s_m$ to the enlargement of the physical space $N$, meant as the number of elements composing the system, Eq.~\eqref{eq:coherence}; we call a system or a model Zipfian if
\[
    \frac{d\log\ton*{\frac{s_M}{s_m}}}{dn}\geq\gamma\frac{d\log N}{dn}
\]
This expression implies that Zipfian systems, i.e. those that satisfy the above inequality, are attracted toward Zipf's law and evolve out of equilibrium, never fully sampling their probabilistic space. Moreover, by generalizing the treatment of Lu et al \cite{lu2010}, we demonstrate that Heaps' law is a particular case of Zipfian dynamics.

Conversely some systems, such as earthquakes, show Zipf's only temporarily. We call this effect, a consequence of a possibly accidental under-sampling, \textit{spurious} Zipf's law. In this case, the growth of $N$ is fast enough to ensure a sampling of the large events, this allowing to estimate the upper cutoff of the PDF. For instance, we determine the maximal possible magnitude of an earthquake occurring in Italy and we show that, on the contrary, the largest tsunami database does not provide a sufficient statistic for inferring the upper cutoff of the distribution. This technique can be easily generalized to other natural or man-provoked hazards, such as hurricanes or terrorist attacks, both claimed to be power-law distributed \cite{corral2010scaling, clauset2007frequency}. We stress that \textit{a spurious Zipf's law can be identified only by performing a dynamical analysis}, and this opens questions about the effective ubiquity of Zipf's scaling. Indeed, many systems where Zipf's law is claimed to be found, such as world cities and earthquakes, are actually evolving towards a high $Q$ configuration, so departing from Zipf's law. 

Then we studied a number of concrete applications of our quantitative framework of analysis:
\begin{itemize}
    \item we show that earthquakes, being essentially independent (in the sense specified before) and characterized by a fixed upper limit, can evolve only incoherently and show Zipf's law spuriously. As aforementioned we used this property for computing the maximal magnitude of an earthquake occurring in Italy;
    \item natural language dynamics is intrinsically Zipfian thanks to the inherent cohrence provided by the grammar rules. Zipfian dynamics, moreover, increases the efficiency of the language and can be directly related to the renowned optimization argument proposed by Mandelbrot;
    \item US metropolitan areas evolved Zipfianly from the Declaration of Independence and the number of different US cities grew according to Heaps's law;
    \item we analitically show that Zipfian dynamics is not additive, confirming this finding also empirically by comparing the evolution of US versus world cities. Moreover, the dynamics of world cities suggests that the largest urban settlements are getting closer to an intrinsic upper limit of population, that we also estimated.
\end{itemize}

Our framework can be directly applied also to theoretical generative models, in particular we considered multiplicative processes and the Yule-Simon model. We find that both processes are characterized by Zipfian dynamics, but only the latter presents an evolution which, albeit qualitatively, reflects the one of the real systems we considered. In particular, using the Zipf's plane, we show that the Yule-Simon process reproduces the trajectory followed by world countries, suggesting the presence of a rich-get-richer mechanism. This dynamical approach thus allows to understand not only if a model reproduces the emergence of Zipf's scaling, but also if it is suitable for describing the evolution of systems toward the Zipf's regime.

Recently, Corominas-Murtra et al. \cite{SSR} introduced sample-space-reducing processes, showing that they produce Zipf's law. We note that the kind of random walks with space restriction introduced in \cite{SSR} can be seen as a very particular case in which the space that the walker can visit at each times reduces coherently with the previous dynamics. However we stress the fact that the mechanism behind the generation of genuine Zipfian dynamics is from one hand much more general and and from the other much more interesting when the space limits enlarge coherently with the dynamics, as in the case for instance of natural language. Indeed only in this case Zipf law emerges as stable self-averaging dynamical feature of the system. For instance the recently proposed Urn Model with Triggering \cite{Loreto}, based on the concept of adjacent possible \cite{Kauffman}, goes in this direction.\\
In table \ref{tab:my_label} we summarize if the various systems and models we analyzed are Zipfian or not, and the the main specific findings our framework provides. Clearly, our approach can be applied to all systems that are claimed to follow Zipf's law.\\

In short, the main points of our work are the following:
\begin{enumerate}
\item Zipf's law should be studied during its evolution: this is the only way to recognize if the system or, better, its dynamics, is truly Zipfian or not;
\item Zipf's law may appear only temporary and, so, spuriously: in particular, any power-law distributed system can show Zipf's law the underlying PDF is under-sampled;
\item if a system shows a spurious Zipf's, then one can estimate the upper cutoff of the generating distribution;
\item if a system is truly Zipfian, then is inherently out of equilibrium, since its the probabilistic space enlarge faster than the physical one;
\item systems showing Heaps' law are a subset of those characterized by a Zipfian dynamics;
\item studying the dynamics allows to determine if a generative model is capable of explaining not only the emergence of Zipf's law, but also the dynamical evolution of systems toward this scaling regime.
\end{enumerate}

Finally we stress that our study opens also a series of questions which should be deeply investigated. Firstly, most of the analysis concerning Zipf's scaling are performed statically, by checking the straightness of the rank-size plot at a given time. As a consequence the list of systems showing genuine Zipf's law may be drastically smaller than usually claimed. Focusing only on genuine Zipfian systems may allow to find a universal generating mechanism for Zipf's law; clearly such a goal is not achievable without the exclusion of spurious Zipfian systems. For example the Zipfian distribution of Lunar craters, a never explained manifestation of this scaling law, could be addressed as the result of an accidental under-sampling provoked by the low rate of asteroids collisions and so as a spurious manifestation of Zipf's law. The possibility of analyzing systems from a never considered dynamical perspective is another novel aspect made available by our framework. For instance we noticed that while some systems, such as natural language or metro areas, present a regular evolution of $Q$, other systems, such as that formed by world conflicts, show a dynamics characterized by sudden decreases of $Q$. This behavior could be explained in terms of a jump of the upper cutoff, as follows from Eq.~\eqref{eq:bar_s_Q}, and therefore our framework can also be used for gathering novel insight on that phenomena usually defined Black Swans. These and other topics will be the object of future studies. 

\section{Methods}
\subsection{Derivation of the relation between the PDF of sizes and the Zipf-Mandelbrot distribution}
Let us consider the truncated power law distribution of sizes, $P(S)$, expressed in Eq.(\ref{trunc}),
where $c$ is the normalization constant, and $s_{m}$ and $s_{M}$ respectively correspond to the natural lower and upper cutoffs, always present in real systems. These cutoffs are connected to $c$ by the normalization condition
\begin{equation}
c\int\limits_{s_m}^{s_M}\frac{ds}{s^{\alpha}}=1 \ \rightarrow \ c=\frac{\alpha-1}{s_m^{1-\alpha}-s_M^{1-\alpha}}
\label{eq:c}
\end{equation}
It is possible to express the rank-size relation as a function of the PDF parameters using the fact that given the PDF $P(S)$ of a continuous variable $S$, the values of its Cumulative Distribution Function (CDF) $C(S)$, associated to the different values of S, are approximately equiprobable. In fact if $P(s)$ is the PDF of the variable $S$ defined in the interval $[s_m,s_M]$, then $C(S)=\int_{s_m}^S ds'\,P(s')$. By performing the change of variables from $S$ to $C=C(S)$, and calling $f(C)$ its PDF, we get by definition of PDF and CDF $f(C)=\frac{dS(C)}{dC}P(S)|_{S=S(C)}=1$ for $0\le C\le 1$. This implies that, given $N$ values of $S$ independently extracted from $P(S)$, with good approximation they can be taken as uniformly spaced in the corresponding variable $C$. Thus, the $k^{\mbox{\small{th}}}$ size ranked value $S(k)$ approximately corresponds to the CDF value  $\frac{N+1-k}{N+1}$. In formulas
        \[
            \int\limits_{s_m}^{S(k)}P(S)dS=c\int\limits_{s_m}^{S(k)}\frac{ds}{s^{\alpha}}\simeq\frac{N+1-k}{N+1}\,,
        \]
       which, together to Eq.~\eqref{eq:c}, gives
        \[
            \frac{S(k)^{1-\alpha}-s_m^{1-\alpha}}{s_M^{1-\alpha}-s_m^{1-\alpha}}\simeq\frac{N+1-k}{N+1} \,.
	    \]
	    By assuming $N+1\approx N$, $s_M\gg s_m$, and introducing $\gamma=\frac{1}{\alpha-1}$, we end up with the final rank-size formula  
        \begin{equation*}
    	    S(k)=\qua*{\frac{Ns_m^{\frac{1}{\gamma}}s_M^{\frac{1}{\gamma}}}{Ns_m^{\frac{1}{\gamma}}+ks_M^{\frac{1}{\gamma}}}}^{\gamma}=\frac{N^{\gamma}s_m}{\qua*{k+N\ton*{\frac{s_m}{s_M}}^{\frac{1}{\gamma}}}^{\gamma}}\,.
        \end{equation*}
A similar computation has been performed by Lu et al. \cite{lu2010} in order to study the relation between Heaps and Zipf's laws. By comparing Eqs.~\eqref{eq:rank_size} and \eqref{eq:Zipf_Mandelbrot} we can derive the following expressions 
        \begin{equation*}
    	    \begin{cases}
    	        \gamma=\frac{1}{\alpha-1}\\
    		    \bar{S}=N^{\gamma}s_m\\
    		    Q=N\ton*{\frac{s_m}{s_M}}^{\frac{1}{\gamma}}\,
    	    \end{cases}
        \end{equation*}
		that relate the number of values$/$objects and the parameters of the PDF $P(S)$ on one side, and the Zipf-Mandelbrot parameters on the other, that is Eq.(\eqref{eq:bar_s_Q}) of the Results section.

\subsection{Databases used}
All the databases we used are freely accessible on the web. In the following we shortly describe them.
\begin{itemize}
\item \textbf{Earthquakes} For our analysis of the Italian earthquakes we used the INGV Parametric Catalogue of Italian Earthquakes, which "provides homogeneous macroseismic and instrumental data and parameters for Italian earthquakes with maximum intensity $\geq5$ or magnitude $geq4.0$ in the period 1000-2017" \citep{Earthquakes}. For what concerns Californian ones we used the California Department of conservation \href{https://www.conservation.ca.gov/cgs/Pages/Earthquakes/earthquake-catalog.aspx}{dataset}. It ranges from 1769 to 2000 and is an extension of Petersen et al. catalog \cite{petersen1996}.
\item \textbf{Tsunami} The study of world tsunami has been performed using NOAA NCEI/WDS Global Historical Tsunami Database, 2100 BC to Present \cite{noaaTsunami}.
\item \textbf{US metropolitan areas} Our study of US metro areas is based on the work of Schroeder \cite{Metro_areas}. The database, which contains historical estimates of US metro areas and counties population, can be accessed \href{https://conservancy.umn.edu/handle/11299/181605}{here}. The number of different US cities and the urban population can be found in \cite{commerce}.
\item \textbf{World cities} The historical population of world cities comes from the World Urbanization Prospects 2018 of the Department of Economic and Social Affairs (UN) \cite{world_cities}. In particular we used the "Annual Population of Urban Agglomerations with 300,000 Inhabitants or More in 2018, by country, 1950-2035" (ignoring 2021-2035 data).
\item \textbf{Language} In order to study the evolution of language we used CHILDES database \cite{macwhinney2000}. In particular we analyzed with CLAN program all the American English corpora available to obtain two rank-size list, one referred to children below six and the other to adults. 
\item \textbf{GDP PPP of countries} Maddison database \cite{maddison2013}, available \href{https://www.rug.nl/ggdc/historicaldevelopment/maddison/releases/maddison-database-2010}{here}, provides GDP PPP of countries from 1 AD to 2008. We integrated it with IMF \href{https://en.wikipedia.org/wiki/List_of_countries_by_past_and_projected_GDP_(PPP)#cite_note-1}{data} to obtain a database which ranges from 1900 to 2019. 
\item \textbf{Volcanic eruptions} The casualties provoked by volcanic eruptions have been collected from NOAA historical dataset, which ranges from 4300 BC to present.
\end{itemize}
\subsection{Fitting procedure}
If an high level of precision is needed, the fitting of power law distributions is a particularly difficult procedure which has been studied extensively \cite{burroughs2001, clauset2009}. In particular, it has been shown that using least squares techniques give biased estimates of the slope of the PDF. However, in our work we are not interested in obtaining a precise estimate of the parameters of the PDF, but rather in checking the trend they follow. We then adopted a standard non linear least squares fitting procedure, whose accuracy, when applied to the rank-size plot or to the complementary cumulative distribution, is comparable to maximum likelihood estimates \cite{white2008estimating}. In particular, we used Eq.~\eqref{eq:rank_size} partially linearized through logarithms
		\[
			\log S(k)=-\frac{1}{\alpha-1}\log\qua*{k+N\ton*{\frac{s_m}{s_M}}^{\alpha-1}}+\log\ton*{N^{\frac{1}{\alpha-1}}s_m}
		\] 
The use of a more sophisticated technique would probably remove some noise from the trajectories, but in our opinion the trend is clear also with the procedure we followed. The presence or absence of deviations at first ranks can be easily checked at glance and is definitely less problematic than the computation of an unbiased estimator of the slope.
	    \begin{acknowledgments}
            We thank Michael Batty for his helpful comments.
        \end{acknowledgments}

\begin{thebibliography}{50}%
\makeatletter
\providecommand \@ifxundefined [1]{%
 \@ifx{#1\undefined}
}%
\providecommand \@ifnum [1]{%
 \ifnum #1\expandafter \@firstoftwo
 \else \expandafter \@secondoftwo
 \fi
}%
\providecommand \@ifx [1]{%
 \ifx #1\expandafter \@firstoftwo
 \else \expandafter \@secondoftwo
 \fi
}%
\providecommand \natexlab [1]{#1}%
\providecommand \enquote  [1]{``#1''}%
\providecommand \bibnamefont  [1]{#1}%
\providecommand \bibfnamefont [1]{#1}%
\providecommand \citenamefont [1]{#1}%
\providecommand \href@noop [0]{\@secondoftwo}%
\providecommand \href [0]{\begingroup \@sanitize@url \@href}%
\providecommand \@href[1]{\@@startlink{#1}\@@href}%
\providecommand \@@href[1]{\endgroup#1\@@endlink}%
\providecommand \@sanitize@url [0]{\catcode `\\12\catcode `\$12\catcode
  `\&12\catcode `\#12\catcode `\^12\catcode `\_12\catcode `\%12\relax}%
\providecommand \@@startlink[1]{}%
\providecommand \@@endlink[0]{}%
\providecommand \url  [0]{\begingroup\@sanitize@url \@url }%
\providecommand \@url [1]{\endgroup\@href {#1}{\urlprefix }}%
\providecommand \urlprefix  [0]{URL }%
\providecommand \Eprint [0]{\href }%
\providecommand \doibase [0]{http://dx.doi.org/}%
\providecommand \selectlanguage [0]{\@gobble}%
\providecommand \bibinfo  [0]{\@secondoftwo}%
\providecommand \bibfield  [0]{\@secondoftwo}%
\providecommand \translation [1]{[#1]}%
\providecommand \BibitemOpen [0]{}%
\providecommand \bibitemStop [0]{}%
\providecommand \bibitemNoStop [0]{.\EOS\space}%
\providecommand \EOS [0]{\spacefactor3000\relax}%
\providecommand \BibitemShut  [1]{\csname bibitem#1\endcsname}%
\let\auto@bib@innerbib\@empty
\bibitem [{\citenamefont {Zipf}(2016)}]{Language}%
  \BibitemOpen
  \bibfield  {author} {\bibinfo {author} {\bibfnamefont {George~Kingsley}\
  \bibnamefont {Zipf}},\ }\href@noop {} {\emph {\bibinfo {title} {Human
  behavior and the principle of least effort: An introduction to human
  ecology}}}\ (\bibinfo  {publisher} {Ravenio Books},\ \bibinfo {year}
  {2016})\BibitemShut {NoStop}%
\bibitem [{\citenamefont {Gell-Mann}(1995)}]{Jaguar}%
  \BibitemOpen
  \bibfield  {author} {\bibinfo {author} {\bibfnamefont {Murray}\ \bibnamefont
  {Gell-Mann}},\ }\href@noop {} {\emph {\bibinfo {title} {The Quark and the
  Jaguar: Adventures in the Simple and the Complex}}}\ (\bibinfo  {publisher}
  {Macmillan},\ \bibinfo {year} {1995})\BibitemShut {NoStop}%
\bibitem [{\citenamefont {Cristelli}\ \emph {et~al.}(2012)\citenamefont
  {Cristelli}, \citenamefont {Batty},\ and\ \citenamefont
  {Pietronero}}]{Batty}%
  \BibitemOpen
  \bibfield  {author} {\bibinfo {author} {\bibfnamefont {Matthieu}\
  \bibnamefont {Cristelli}}, \bibinfo {author} {\bibfnamefont {Michael}\
  \bibnamefont {Batty}}, \ and\ \bibinfo {author} {\bibfnamefont {Luciano}\
  \bibnamefont {Pietronero}},\ }\bibfield  {title} {\enquote {\bibinfo {title}
  {There is more than a power law in zipf},}\ }\href@noop {} {\bibfield
  {journal} {\bibinfo  {journal} {Scientific reports}\ }\textbf {\bibinfo
  {volume} {2}},\ \bibinfo {pages} {812} (\bibinfo {year} {2012})}\BibitemShut
  {NoStop}%
\bibitem [{\citenamefont {Axtell}(2001)}]{Firms}%
  \BibitemOpen
  \bibfield  {author} {\bibinfo {author} {\bibfnamefont {Robert~L}\
  \bibnamefont {Axtell}},\ }\bibfield  {title} {\enquote {\bibinfo {title}
  {Zipf distribution of us firm sizes},}\ }\href@noop {} {\bibfield  {journal}
  {\bibinfo  {journal} {Science}\ }\textbf {\bibinfo {volume} {293}},\ \bibinfo
  {pages} {1818--1820} (\bibinfo {year} {2001})}\BibitemShut {NoStop}%
\bibitem [{\citenamefont {Cunha}\ \emph {et~al.}(1995)\citenamefont {Cunha},
  \citenamefont {Bestavros},\ and\ \citenamefont {Crovella}}]{Web}%
  \BibitemOpen
  \bibfield  {author} {\bibinfo {author} {\bibfnamefont {Carlos~R}\
  \bibnamefont {Cunha}}, \bibinfo {author} {\bibfnamefont {Azer}\ \bibnamefont
  {Bestavros}}, \ and\ \bibinfo {author} {\bibfnamefont {Mark~E}\ \bibnamefont
  {Crovella}},\ }\href@noop {} {\emph {\bibinfo {title} {Characteristics of WWW
  client-based traces}}},\ \bibinfo {type} {Tech. Rep.}\ (\bibinfo
  {institution} {Boston University Computer Science Department},\ \bibinfo
  {year} {1995})\BibitemShut {NoStop}%
\bibitem [{\citenamefont {Redner}(1998)}]{Citations}%
  \BibitemOpen
  \bibfield  {author} {\bibinfo {author} {\bibfnamefont {Sidney}\ \bibnamefont
  {Redner}},\ }\bibfield  {title} {\enquote {\bibinfo {title} {How popular is
  your paper? an empirical study of the citation distribution},}\ }\href@noop
  {} {\bibfield  {journal} {\bibinfo  {journal} {The European Physical Journal
  B-Condensed Matter and Complex Systems}\ }\textbf {\bibinfo {volume} {4}},\
  \bibinfo {pages} {131--134} (\bibinfo {year} {1998})}\BibitemShut {NoStop}%
\bibitem [{\citenamefont {Newman}(2005)}]{Newman}%
  \BibitemOpen
  \bibfield  {author} {\bibinfo {author} {\bibfnamefont {Mark~EJ}\ \bibnamefont
  {Newman}},\ }\bibfield  {title} {\enquote {\bibinfo {title} {Power laws,
  pareto distributions and zipf's law},}\ }\href@noop {} {\bibfield  {journal}
  {\bibinfo  {journal} {Contemporary physics}\ }\textbf {\bibinfo {volume}
  {46}},\ \bibinfo {pages} {323--351} (\bibinfo {year} {2005})}\BibitemShut
  {NoStop}%
\bibitem [{\citenamefont {Sornette}\ \emph {et~al.}(1996)\citenamefont
  {Sornette}, \citenamefont {Knopoff}, \citenamefont {Kagan},\ and\
  \citenamefont {Vanneste}}]{sornette1996rank}%
  \BibitemOpen
  \bibfield  {author} {\bibinfo {author} {\bibfnamefont {Didier}\ \bibnamefont
  {Sornette}}, \bibinfo {author} {\bibfnamefont {Leon}\ \bibnamefont
  {Knopoff}}, \bibinfo {author} {\bibfnamefont {YY}~\bibnamefont {Kagan}}, \
  and\ \bibinfo {author} {\bibfnamefont {Christian}\ \bibnamefont {Vanneste}},\
  }\bibfield  {title} {\enquote {\bibinfo {title} {Rank-ordering statistics of
  extreme events: Application to the distribution of large earthquakes},}\
  }\href@noop {} {\bibfield  {journal} {\bibinfo  {journal} {Journal of
  Geophysical Research: Solid Earth}\ }\textbf {\bibinfo {volume} {101}},\
  \bibinfo {pages} {13883--13893} (\bibinfo {year} {1996})}\BibitemShut
  {NoStop}%
\bibitem [{\citenamefont {Pietronero}\ \emph {et~al.}(2001)\citenamefont
  {Pietronero}, \citenamefont {Tosatti}, \citenamefont {Tosatti},\ and\
  \citenamefont {Vespignani}}]{Multiplicative}%
  \BibitemOpen
  \bibfield  {author} {\bibinfo {author} {\bibfnamefont {Luciano}\ \bibnamefont
  {Pietronero}}, \bibinfo {author} {\bibfnamefont {Erio}\ \bibnamefont
  {Tosatti}}, \bibinfo {author} {\bibfnamefont {Valentino}\ \bibnamefont
  {Tosatti}}, \ and\ \bibinfo {author} {\bibfnamefont {Alessandro}\
  \bibnamefont {Vespignani}},\ }\bibfield  {title} {\enquote {\bibinfo {title}
  {Explaining the uneven distribution of numbers in nature: the laws of benford
  and zipf},}\ }\href@noop {} {\bibfield  {journal} {\bibinfo  {journal}
  {Physica A: Statistical Mechanics and its Applications}\ }\textbf {\bibinfo
  {volume} {293}},\ \bibinfo {pages} {297--304} (\bibinfo {year}
  {2001})}\BibitemShut {NoStop}%
\bibitem [{\citenamefont {Levy}\ and\ \citenamefont
  {Solomon}(1996)}]{levy1996power}%
  \BibitemOpen
  \bibfield  {author} {\bibinfo {author} {\bibfnamefont {Moshe}\ \bibnamefont
  {Levy}}\ and\ \bibinfo {author} {\bibfnamefont {Sorin}\ \bibnamefont
  {Solomon}},\ }\bibfield  {title} {\enquote {\bibinfo {title} {Power laws are
  logarithmic boltzmann laws},}\ }\href@noop {} {\bibfield  {journal} {\bibinfo
   {journal} {International Journal of Modern Physics C}\ }\textbf {\bibinfo
  {volume} {7}},\ \bibinfo {pages} {595--601} (\bibinfo {year}
  {1996})}\BibitemShut {NoStop}%
\bibitem [{\citenamefont {Tria}\ \emph {et~al.}(2014)\citenamefont {Tria},
  \citenamefont {Loreto}, \citenamefont {Servedio},\ and\ \citenamefont
  {Strogatz}}]{Loreto}%
  \BibitemOpen
  \bibfield  {author} {\bibinfo {author} {\bibfnamefont {Francesca}\
  \bibnamefont {Tria}}, \bibinfo {author} {\bibfnamefont {Vittorio}\
  \bibnamefont {Loreto}}, \bibinfo {author} {\bibfnamefont {Vito
  Domenico~Pietro}\ \bibnamefont {Servedio}}, \ and\ \bibinfo {author}
  {\bibfnamefont {Steven~H}\ \bibnamefont {Strogatz}},\ }\bibfield  {title}
  {\enquote {\bibinfo {title} {The dynamics of correlated novelties},}\
  }\href@noop {} {\bibfield  {journal} {\bibinfo  {journal} {Scientific
  reports}\ }\textbf {\bibinfo {volume} {4}},\ \bibinfo {pages} {5890}
  (\bibinfo {year} {2014})}\BibitemShut {NoStop}%
\bibitem [{\citenamefont {Kauffman}(1996)}]{Kauffman}%
  \BibitemOpen
  \bibfield  {author} {\bibinfo {author} {\bibfnamefont {Stuart}\ \bibnamefont
  {Kauffman}},\ }\href@noop {} {\emph {\bibinfo {title} {At home in the
  universe: The search for the laws of self-organization and complexity}}}\
  (\bibinfo  {publisher} {Oxford university press},\ \bibinfo {year}
  {1996})\BibitemShut {NoStop}%
\bibitem [{\citenamefont {Corominas-Murtra}\ \emph {et~al.}(2015)\citenamefont
  {Corominas-Murtra}, \citenamefont {Hanel},\ and\ \citenamefont
  {Thurner}}]{SSR}%
  \BibitemOpen
  \bibfield  {author} {\bibinfo {author} {\bibfnamefont {Bernat}\ \bibnamefont
  {Corominas-Murtra}}, \bibinfo {author} {\bibfnamefont {Rudolf}\ \bibnamefont
  {Hanel}}, \ and\ \bibinfo {author} {\bibfnamefont {Stefan}\ \bibnamefont
  {Thurner}},\ }\bibfield  {title} {\enquote {\bibinfo {title} {Understanding
  scaling through history-dependent processes with collapsing sample space},}\
  }\href@noop {} {\bibfield  {journal} {\bibinfo  {journal} {Proceedings of the
  National Academy of Sciences}\ }\textbf {\bibinfo {volume} {112}},\ \bibinfo
  {pages} {5348--5353} (\bibinfo {year} {2015})}\BibitemShut {NoStop}%
\bibitem [{\citenamefont {Cubero}\ \emph {et~al.}(2019)\citenamefont {Cubero},
  \citenamefont {Jo}, \citenamefont {Marsili}, \citenamefont {Roudi},\ and\
  \citenamefont {Song}}]{Marsili}%
  \BibitemOpen
  \bibfield  {author} {\bibinfo {author} {\bibfnamefont {Ryan~John}\
  \bibnamefont {Cubero}}, \bibinfo {author} {\bibfnamefont {Junghyo}\
  \bibnamefont {Jo}}, \bibinfo {author} {\bibfnamefont {Matteo}\ \bibnamefont
  {Marsili}}, \bibinfo {author} {\bibfnamefont {Yasser}\ \bibnamefont {Roudi}},
  \ and\ \bibinfo {author} {\bibfnamefont {Juyong}\ \bibnamefont {Song}},\
  }\bibfield  {title} {\enquote {\bibinfo {title} {Statistical criticality
  arises in most informative representations},}\ }\href@noop {} {\bibfield
  {journal} {\bibinfo  {journal} {Journal of Statistical Mechanics: Theory and
  Experiment}\ }\textbf {\bibinfo {volume} {2019}},\ \bibinfo {pages} {063402}
  (\bibinfo {year} {2019})}\BibitemShut {NoStop}%
\bibitem [{\citenamefont {Mandelbrot}(1953)}]{Mandelbrot}%
  \BibitemOpen
  \bibfield  {author} {\bibinfo {author} {\bibfnamefont {Benoit}\ \bibnamefont
  {Mandelbrot}},\ }\bibfield  {title} {\enquote {\bibinfo {title} {An
  informational theory of the statistical structure of language},}\ }\href@noop
  {} {\bibfield  {journal} {\bibinfo  {journal} {Communication theory}\
  }\textbf {\bibinfo {volume} {84}},\ \bibinfo {pages} {486--502} (\bibinfo
  {year} {1953})}\BibitemShut {NoStop}%
\bibitem [{\citenamefont {Yule}(1925)}]{yule1925ii}%
  \BibitemOpen
  \bibfield  {author} {\bibinfo {author} {\bibfnamefont {George~Udny}\
  \bibnamefont {Yule}},\ }\bibfield  {title} {\enquote {\bibinfo {title}
  {Ii.—a mathematical theory of evolution, based on the conclusions of dr. jc
  willis, fr s},}\ }\href@noop {} {\bibfield  {journal} {\bibinfo  {journal}
  {Philosophical transactions of the Royal Society of London. Series B,
  containing papers of a biological character}\ }\textbf {\bibinfo {volume}
  {213}},\ \bibinfo {pages} {21--87} (\bibinfo {year} {1925})}\BibitemShut
  {NoStop}%
\bibitem [{\citenamefont {Simon}(1955)}]{simon1955class}%
  \BibitemOpen
  \bibfield  {author} {\bibinfo {author} {\bibfnamefont {Herbert~A}\
  \bibnamefont {Simon}},\ }\bibfield  {title} {\enquote {\bibinfo {title} {On a
  class of skew distribution functions},}\ }\href@noop {} {\bibfield  {journal}
  {\bibinfo  {journal} {Biometrika}\ }\textbf {\bibinfo {volume} {42}},\
  \bibinfo {pages} {425--440} (\bibinfo {year} {1955})}\BibitemShut {NoStop}%
\bibitem [{\citenamefont {Heaps}(1978)}]{heaps1978information}%
  \BibitemOpen
  \bibfield  {author} {\bibinfo {author} {\bibfnamefont {Harold~Stanley}\
  \bibnamefont {Heaps}},\ }\href@noop {} {\emph {\bibinfo {title} {Information
  retrieval, computational and theoretical aspects}}}\ (\bibinfo  {publisher}
  {Academic Press},\ \bibinfo {year} {1978})\BibitemShut {NoStop}%
\bibitem [{\citenamefont {L{\"u}}\ \emph {et~al.}(2010)\citenamefont {L{\"u}},
  \citenamefont {Zhang},\ and\ \citenamefont {Zhou}}]{lu2010}%
  \BibitemOpen
  \bibfield  {author} {\bibinfo {author} {\bibfnamefont {Linyuan}\ \bibnamefont
  {L{\"u}}}, \bibinfo {author} {\bibfnamefont {Zi-Ke}\ \bibnamefont {Zhang}}, \
  and\ \bibinfo {author} {\bibfnamefont {Tao}\ \bibnamefont {Zhou}},\
  }\bibfield  {title} {\enquote {\bibinfo {title} {Zipf's law leads to heaps'
  law: Analyzing their relation in finite-size systems},}\ }\href@noop {}
  {\bibfield  {journal} {\bibinfo  {journal} {PloS one}\ }\textbf {\bibinfo
  {volume} {5}},\ \bibinfo {pages} {e14139} (\bibinfo {year}
  {2010})}\BibitemShut {NoStop}%
\bibitem [{Note1()}]{Note1}%
  \BibitemOpen
  \bibinfo {note} {The effect of the only upper cutoffs has been previously
  considered \cite {burroughs2001, burroughs2001b}; in any case, never from a
  dynamical point of view.}\BibitemShut {Stop}%
\bibitem [{\citenamefont {Li}(2002)}]{li2002}%
  \BibitemOpen
  \bibfield  {author} {\bibinfo {author} {\bibfnamefont {Wentian}\ \bibnamefont
  {Li}},\ }\bibfield  {title} {\enquote {\bibinfo {title} {Zipf's law
  everywhere.}}\ }\href@noop {} {\bibfield  {journal} {\bibinfo  {journal}
  {Glottometrics}\ }\textbf {\bibinfo {volume} {5}},\ \bibinfo {pages} {14--21}
  (\bibinfo {year} {2002})}\BibitemShut {NoStop}%
\bibitem [{\citenamefont {Rovida}\ \emph {et~al.}(2016)\citenamefont {Rovida},
  \citenamefont {Locati}, \citenamefont {Camassi}, \citenamefont {Lolli},\ and\
  \citenamefont {Gasperini}}]{Earthquakes}%
  \BibitemOpen
  \bibfield  {author} {\bibinfo {author} {\bibfnamefont {Andrea~Nicola}\
  \bibnamefont {Rovida}}, \bibinfo {author} {\bibfnamefont {Mario}\
  \bibnamefont {Locati}}, \bibinfo {author} {\bibfnamefont {Romano~Daniele}\
  \bibnamefont {Camassi}}, \bibinfo {author} {\bibfnamefont {Barbara}\
  \bibnamefont {Lolli}}, \ and\ \bibinfo {author} {\bibfnamefont {Paolo}\
  \bibnamefont {Gasperini}},\ }\bibfield  {title} {\enquote {\bibinfo {title}
  {Cpti15, the 2015 version of the parametric catalogue of italian
  earthquakes},}\ }\href@noop {} {\  (\bibinfo {year} {2016})}\BibitemShut
  {NoStop}%
\bibitem [{\citenamefont {Schroeder}(2016)}]{Metro_areas}%
  \BibitemOpen
  \bibfield  {author} {\bibinfo {author} {\bibfnamefont {Jonathan~P}\
  \bibnamefont {Schroeder}},\ }\bibfield  {title} {\enquote {\bibinfo {title}
  {Historical population estimates for 2010 us states, counties and metro/micro
  areas, 1790-2010},}\ }\href@noop {} {\bibfield  {journal} {\bibinfo
  {journal} {Retrieved from the Data Repository for the University of
  Minnesota}\ } (\bibinfo {year} {2016})}\BibitemShut {NoStop}%
\bibitem [{\citenamefont {Center}(2020{\natexlab{a}})}]{noaaVolcanoes}%
  \BibitemOpen
  \bibfield  {author} {\bibinfo {author} {\bibfnamefont {National
  Geophysical~Data}\ \bibnamefont {Center}},\ }\bibfield  {title} {\enquote
  {\bibinfo {title} {World data service: Ncei/wds global significant volcanic
  eruptions database},}\ }\href {\doibase 10.7289/V5JW8BSH} {\  (\bibinfo
  {year} {2020}{\natexlab{a}}),\ 10.7289/V5JW8BSH}\BibitemShut {NoStop}%
\bibitem [{\citenamefont {Mazzolini}\ \emph {et~al.}(2018)\citenamefont
  {Mazzolini}, \citenamefont {Colliva}, \citenamefont {Caselle},\ and\
  \citenamefont {Osella}}]{mazzolini2018heaps}%
  \BibitemOpen
  \bibfield  {author} {\bibinfo {author} {\bibfnamefont {Andrea}\ \bibnamefont
  {Mazzolini}}, \bibinfo {author} {\bibfnamefont {Alberto}\ \bibnamefont
  {Colliva}}, \bibinfo {author} {\bibfnamefont {Michele}\ \bibnamefont
  {Caselle}}, \ and\ \bibinfo {author} {\bibfnamefont {Matteo}\ \bibnamefont
  {Osella}},\ }\bibfield  {title} {\enquote {\bibinfo {title} {Heaps' law,
  statistics of shared components, and temporal patterns from a
  sample-space-reducing process},}\ }\href@noop {} {\bibfield  {journal}
  {\bibinfo  {journal} {Physical Review E}\ }\textbf {\bibinfo {volume} {98}},\
  \bibinfo {pages} {052139} (\bibinfo {year} {2018})}\BibitemShut {NoStop}%
\bibitem [{\citenamefont {Gutenberg}\ and\ \citenamefont
  {Richter}(1936)}]{Gutenberg}%
  \BibitemOpen
  \bibfield  {author} {\bibinfo {author} {\bibfnamefont {Beno}\ \bibnamefont
  {Gutenberg}}\ and\ \bibinfo {author} {\bibfnamefont {Charles~Francis}\
  \bibnamefont {Richter}},\ }\bibfield  {title} {\enquote {\bibinfo {title}
  {Magnitude and energy of earthquakes},}\ }\href@noop {} {\bibfield  {journal}
  {\bibinfo  {journal} {Science}\ }\textbf {\bibinfo {volume} {83}},\ \bibinfo
  {pages} {183--185} (\bibinfo {year} {1936})}\BibitemShut {NoStop}%
\bibitem [{\citenamefont {Center}(2020{\natexlab{b}})}]{noaaTsunami}%
  \BibitemOpen
  \bibfield  {author} {\bibinfo {author} {\bibfnamefont {National
  Geophysical~Data}\ \bibnamefont {Center}},\ }\bibfield  {title} {\enquote
  {\bibinfo {title} {World data service: Ncei/wds global historical tsunami
  database},}\ }\href {\doibase 10.7289/V5PN93H7} {\  (\bibinfo {year}
  {2020}{\natexlab{b}}),\ 10.7289/V5PN93H7}\BibitemShut {NoStop}%
\bibitem [{Note2()}]{Note2}%
  \BibitemOpen
  \bibinfo {note} {Note that the size of the smallest urban settlement, $s_m$,
  is by a good extent constant: even if existing cities grow, new ones are
  constantly founded, avoiding the growth of the lower cutoff.}\BibitemShut
  {Stop}%
\bibitem [{\citenamefont {Simini}\ and\ \citenamefont {James}(2019)}]{simini}%
  \BibitemOpen
  \bibfield  {author} {\bibinfo {author} {\bibfnamefont {Filippo}\ \bibnamefont
  {Simini}}\ and\ \bibinfo {author} {\bibfnamefont {Charlotte}\ \bibnamefont
  {James}},\ }\bibfield  {title} {\enquote {\bibinfo {title} {Testing heaps’
  law for cities using administrative and gridded population data sets},}\
  }\href@noop {} {\bibfield  {journal} {\bibinfo  {journal} {EPJ Data Science}\
  }\textbf {\bibinfo {volume} {8}},\ \bibinfo {pages} {1--13} (\bibinfo {year}
  {2019})}\BibitemShut {NoStop}%
\bibitem [{\citenamefont {Soo}(2005)}]{soo2005zipf}%
  \BibitemOpen
  \bibfield  {author} {\bibinfo {author} {\bibfnamefont {Kwok~Tong}\
  \bibnamefont {Soo}},\ }\bibfield  {title} {\enquote {\bibinfo {title} {Zipf's
  law for cities: a cross-country investigation},}\ }\href@noop {} {\bibfield
  {journal} {\bibinfo  {journal} {Regional science and urban Economics}\
  }\textbf {\bibinfo {volume} {35}},\ \bibinfo {pages} {239--263} (\bibinfo
  {year} {2005})}\BibitemShut {NoStop}%
\bibitem [{\citenamefont {United~Nations}\ and\ \citenamefont
  {Social~Affairs}(2018)}]{world_cities}%
  \BibitemOpen
  \bibfield  {author} {\bibinfo {author} {\bibfnamefont {Department
  of~Economic}\ \bibnamefont {United~Nations}}\ and\ \bibinfo {author}
  {\bibfnamefont {Population~Division}\ \bibnamefont {Social~Affairs}},\
  }\href@noop {} {\enquote {\bibinfo {title} {World urbanization prospects: The
  2018 revision},}\ } (\bibinfo {year} {2018})\BibitemShut {NoStop}%
\bibitem [{\citenamefont {Molina}\ and\ \citenamefont
  {Molina}(2004)}]{molina2004megacities}%
  \BibitemOpen
  \bibfield  {author} {\bibinfo {author} {\bibfnamefont {Mario~J}\ \bibnamefont
  {Molina}}\ and\ \bibinfo {author} {\bibfnamefont {Luisa~T}\ \bibnamefont
  {Molina}},\ }\bibfield  {title} {\enquote {\bibinfo {title} {Megacities and
  atmospheric pollution},}\ }\href@noop {} {\bibfield  {journal} {\bibinfo
  {journal} {Journal of the Air \& Waste Management Association}\ }\textbf
  {\bibinfo {volume} {54}},\ \bibinfo {pages} {644--680} (\bibinfo {year}
  {2004})}\BibitemShut {NoStop}%
\bibitem [{\citenamefont {Varis}\ \emph {et~al.}(2006)\citenamefont {Varis},
  \citenamefont {Biswas}, \citenamefont {Tortajada},\ and\ \citenamefont
  {Lundqvist}}]{varis2006megacities}%
  \BibitemOpen
  \bibfield  {author} {\bibinfo {author} {\bibfnamefont {Olli}\ \bibnamefont
  {Varis}}, \bibinfo {author} {\bibfnamefont {Asit~K}\ \bibnamefont {Biswas}},
  \bibinfo {author} {\bibfnamefont {Cecilia}\ \bibnamefont {Tortajada}}, \ and\
  \bibinfo {author} {\bibfnamefont {Jan}\ \bibnamefont {Lundqvist}},\
  }\bibfield  {title} {\enquote {\bibinfo {title} {Megacities and water
  management},}\ }\href@noop {} {\bibfield  {journal} {\bibinfo  {journal}
  {Water Resources Development}\ }\textbf {\bibinfo {volume} {22}},\ \bibinfo
  {pages} {377--394} (\bibinfo {year} {2006})}\BibitemShut {NoStop}%
\bibitem [{\citenamefont {Wenzel}\ \emph {et~al.}(2007)\citenamefont {Wenzel},
  \citenamefont {Bendimerad},\ and\ \citenamefont
  {Sinha}}]{wenzel2007megacities}%
  \BibitemOpen
  \bibfield  {author} {\bibinfo {author} {\bibfnamefont {Friedemann}\
  \bibnamefont {Wenzel}}, \bibinfo {author} {\bibfnamefont {Fouad}\
  \bibnamefont {Bendimerad}}, \ and\ \bibinfo {author} {\bibfnamefont {Ravi}\
  \bibnamefont {Sinha}},\ }\bibfield  {title} {\enquote {\bibinfo {title}
  {Megacities--megarisks},}\ }\href@noop {} {\bibfield  {journal} {\bibinfo
  {journal} {Natural Hazards}\ }\textbf {\bibinfo {volume} {42}},\ \bibinfo
  {pages} {481--491} (\bibinfo {year} {2007})}\BibitemShut {NoStop}%
\bibitem [{\citenamefont {Hoornweg}\ and\ \citenamefont
  {Pope}(2017)}]{hoornweg2017population}%
  \BibitemOpen
  \bibfield  {author} {\bibinfo {author} {\bibfnamefont {Daniel}\ \bibnamefont
  {Hoornweg}}\ and\ \bibinfo {author} {\bibfnamefont {Kevin}\ \bibnamefont
  {Pope}},\ }\bibfield  {title} {\enquote {\bibinfo {title} {Population
  predictions for the world’s largest cities in the 21st century},}\
  }\href@noop {} {\bibfield  {journal} {\bibinfo  {journal} {Environment and
  Urbanization}\ }\textbf {\bibinfo {volume} {29}},\ \bibinfo {pages}
  {195--216} (\bibinfo {year} {2017})}\BibitemShut {NoStop}%
\bibitem [{\citenamefont {MacWhinney}(2000)}]{macwhinney2000}%
  \BibitemOpen
  \bibfield  {author} {\bibinfo {author} {\bibfnamefont {Brian}\ \bibnamefont
  {MacWhinney}},\ }\href@noop {} {\enquote {\bibinfo {title} {The childes
  project: Tools for analyzing talk: Volume i: Transcription format and
  programs, volume ii: The database},}\ } (\bibinfo {year} {2000})\BibitemShut
  {NoStop}%
\bibitem [{\citenamefont {Mitzenmacher}(2004)}]{mitzenmacher2004brief}%
  \BibitemOpen
  \bibfield  {author} {\bibinfo {author} {\bibfnamefont {Michael}\ \bibnamefont
  {Mitzenmacher}},\ }\bibfield  {title} {\enquote {\bibinfo {title} {A brief
  history of generative models for power law and lognormal distributions},}\
  }\href@noop {} {\bibfield  {journal} {\bibinfo  {journal} {Internet
  mathematics}\ }\textbf {\bibinfo {volume} {1}},\ \bibinfo {pages} {226--251}
  (\bibinfo {year} {2004})}\BibitemShut {NoStop}%
\bibitem [{\citenamefont {Manin}(2009)}]{manin2009mandelbrot}%
  \BibitemOpen
  \bibfield  {author} {\bibinfo {author} {\bibfnamefont {D~Yu}\ \bibnamefont
  {Manin}},\ }\bibfield  {title} {\enquote {\bibinfo {title} {Mandelbrot's
  model for zipf's law: Can mandelbrot's model explain zipf's law for
  language?}}\ }\href@noop {} {\bibfield  {journal} {\bibinfo  {journal}
  {Journal of Quantitative Linguistics}\ }\textbf {\bibinfo {volume} {16}},\
  \bibinfo {pages} {274--285} (\bibinfo {year} {2009})}\BibitemShut {NoStop}%
\bibitem [{\citenamefont {Redner}(1990)}]{redner1990random}%
  \BibitemOpen
  \bibfield  {author} {\bibinfo {author} {\bibfnamefont {Sidney}\ \bibnamefont
  {Redner}},\ }\bibfield  {title} {\enquote {\bibinfo {title} {Random
  multiplicative processes: An elementary tutorial},}\ }\href@noop {}
  {\bibfield  {journal} {\bibinfo  {journal} {American Journal of Physics}\
  }\textbf {\bibinfo {volume} {58}},\ \bibinfo {pages} {267--273} (\bibinfo
  {year} {1990})}\BibitemShut {NoStop}%
\bibitem [{\citenamefont {Sornette}\ and\ \citenamefont
  {Cont}(1997)}]{sornette1997convergent}%
  \BibitemOpen
  \bibfield  {author} {\bibinfo {author} {\bibfnamefont {Didier}\ \bibnamefont
  {Sornette}}\ and\ \bibinfo {author} {\bibfnamefont {Rama}\ \bibnamefont
  {Cont}},\ }\bibfield  {title} {\enquote {\bibinfo {title} {Convergent
  multiplicative processes repelled from zero: power laws and truncated power
  laws},}\ }\href@noop {} {\bibfield  {journal} {\bibinfo  {journal} {Journal
  de Physique I}\ }\textbf {\bibinfo {volume} {7}},\ \bibinfo {pages}
  {431--444} (\bibinfo {year} {1997})}\BibitemShut {NoStop}%
\bibitem [{Note3()}]{Note3}%
  \BibitemOpen
  \bibinfo {note} {We recall that $n=\DOTSB \sum@ \slimits@
  _{k=1}^NS(k)$.}\BibitemShut {Stop}%
\bibitem [{\citenamefont {Corral}\ \emph {et~al.}(2010)\citenamefont {Corral},
  \citenamefont {Oss{\'o}},\ and\ \citenamefont {Llebot}}]{corral2010scaling}%
  \BibitemOpen
  \bibfield  {author} {\bibinfo {author} {\bibfnamefont {{\'A}lvaro}\
  \bibnamefont {Corral}}, \bibinfo {author} {\bibfnamefont {Albert}\
  \bibnamefont {Oss{\'o}}}, \ and\ \bibinfo {author} {\bibfnamefont
  {Josep~Enric}\ \bibnamefont {Llebot}},\ }\bibfield  {title} {\enquote
  {\bibinfo {title} {Scaling of tropical-cyclone dissipation},}\ }\href@noop {}
  {\bibfield  {journal} {\bibinfo  {journal} {Nature Physics}\ }\textbf
  {\bibinfo {volume} {6}},\ \bibinfo {pages} {693--696} (\bibinfo {year}
  {2010})}\BibitemShut {NoStop}%
\bibitem [{\citenamefont {Clauset}\ \emph {et~al.}(2007)\citenamefont
  {Clauset}, \citenamefont {Young},\ and\ \citenamefont
  {Gleditsch}}]{clauset2007frequency}%
  \BibitemOpen
  \bibfield  {author} {\bibinfo {author} {\bibfnamefont {Aaron}\ \bibnamefont
  {Clauset}}, \bibinfo {author} {\bibfnamefont {Maxwell}\ \bibnamefont
  {Young}}, \ and\ \bibinfo {author} {\bibfnamefont {Kristian~Skrede}\
  \bibnamefont {Gleditsch}},\ }\bibfield  {title} {\enquote {\bibinfo {title}
  {On the frequency of severe terrorist events},}\ }\href@noop {} {\bibfield
  {journal} {\bibinfo  {journal} {Journal of Conflict Resolution}\ }\textbf
  {\bibinfo {volume} {51}},\ \bibinfo {pages} {58--87} (\bibinfo {year}
  {2007})}\BibitemShut {NoStop}%
\bibitem [{\citenamefont {Petersen}(1996)}]{petersen1996}%
  \BibitemOpen
  \bibfield  {author} {\bibinfo {author} {\bibfnamefont {Mark~D}\ \bibnamefont
  {Petersen}},\ }\href@noop {} {\emph {\bibinfo {title} {Probabilistic seismic
  hazard assessment for the state of California}}},\ Vol.~\bibinfo {volume}
  {96}\ (\bibinfo  {publisher} {California Department of Conservation Division
  of Mines and Geology},\ \bibinfo {year} {1996})\BibitemShut {NoStop}%
\bibitem [{\citenamefont {Commerce}(1975)}]{commerce}%
  \BibitemOpen
  \bibfield  {author} {\bibinfo {author} {\bibfnamefont {US}~\bibnamefont
  {Commerce}},\ }\bibfield  {title} {\enquote {\bibinfo {title} {Department of
  commerce, bureau of the census, historical statistics of the united states:
  Colonial times to 1970, bicentennial ed},}\ }\href@noop {} {\bibfield
  {journal} {\bibinfo  {journal} {Washington, DC: US Government Printing
  Office}\ } (\bibinfo {year} {1975})}\BibitemShut {NoStop}%
\bibitem [{\citenamefont {Maddison}(2013)}]{maddison2013}%
  \BibitemOpen
  \bibfield  {author} {\bibinfo {author} {\bibfnamefont {Angus}\ \bibnamefont
  {Maddison}},\ }\bibfield  {title} {\enquote {\bibinfo {title} {The
  maddison-project},}\ }\href@noop {} {\bibfield  {journal} {\bibinfo
  {journal} {l{\'\i}nea] http://www. ggdc. net/maddison/maddison--project/home.
  htm}\ }\textbf {\bibinfo {volume} {1}},\ \bibinfo {pages} {14} (\bibinfo
  {year} {2013})}\BibitemShut {NoStop}%
\bibitem [{\citenamefont {Burroughs}\ and\ \citenamefont
  {Tebbens}(2001{\natexlab{a}})}]{burroughs2001}%
  \BibitemOpen
  \bibfield  {author} {\bibinfo {author} {\bibfnamefont {Stephen~M}\
  \bibnamefont {Burroughs}}\ and\ \bibinfo {author} {\bibfnamefont {SARAH~F}\
  \bibnamefont {Tebbens}},\ }\bibfield  {title} {\enquote {\bibinfo {title}
  {Upper-truncated power laws in natural systems},}\ }\href@noop {} {\bibfield
  {journal} {\bibinfo  {journal} {Pure and Applied Geophysics}\ }\textbf
  {\bibinfo {volume} {158}},\ \bibinfo {pages} {741--757} (\bibinfo {year}
  {2001}{\natexlab{a}})}\BibitemShut {NoStop}%
\bibitem [{\citenamefont {Clauset}\ \emph {et~al.}(2009)\citenamefont
  {Clauset}, \citenamefont {Shalizi},\ and\ \citenamefont
  {Newman}}]{clauset2009}%
  \BibitemOpen
  \bibfield  {author} {\bibinfo {author} {\bibfnamefont {Aaron}\ \bibnamefont
  {Clauset}}, \bibinfo {author} {\bibfnamefont {Cosma~Rohilla}\ \bibnamefont
  {Shalizi}}, \ and\ \bibinfo {author} {\bibfnamefont {Mark~EJ}\ \bibnamefont
  {Newman}},\ }\bibfield  {title} {\enquote {\bibinfo {title} {Power-law
  distributions in empirical data},}\ }\href@noop {} {\bibfield  {journal}
  {\bibinfo  {journal} {SIAM review}\ }\textbf {\bibinfo {volume} {51}},\
  \bibinfo {pages} {661--703} (\bibinfo {year} {2009})}\BibitemShut {NoStop}%
\bibitem [{\citenamefont {White}\ \emph {et~al.}(2008)\citenamefont {White},
  \citenamefont {Enquist},\ and\ \citenamefont {Green}}]{white2008estimating}%
  \BibitemOpen
  \bibfield  {author} {\bibinfo {author} {\bibfnamefont {Ethan~P}\ \bibnamefont
  {White}}, \bibinfo {author} {\bibfnamefont {Brian~J}\ \bibnamefont
  {Enquist}}, \ and\ \bibinfo {author} {\bibfnamefont {Jessica~L}\ \bibnamefont
  {Green}},\ }\bibfield  {title} {\enquote {\bibinfo {title} {On estimating the
  exponent of power-law frequency distributions},}\ }\href@noop {} {\bibfield
  {journal} {\bibinfo  {journal} {Ecology}\ }\textbf {\bibinfo {volume} {89}},\
  \bibinfo {pages} {905--912} (\bibinfo {year} {2008})}\BibitemShut {NoStop}%
\bibitem [{\citenamefont {Burroughs}\ and\ \citenamefont
  {Tebbens}(2001{\natexlab{b}})}]{burroughs2001b}%
  \BibitemOpen
  \bibfield  {author} {\bibinfo {author} {\bibfnamefont {Stephen~M}\
  \bibnamefont {Burroughs}}\ and\ \bibinfo {author} {\bibfnamefont {Sarah~F}\
  \bibnamefont {Tebbens}},\ }\bibfield  {title} {\enquote {\bibinfo {title}
  {Upper-truncated power law distributions},}\ }\href@noop {} {\bibfield
  {journal} {\bibinfo  {journal} {Fractals}\ }\textbf {\bibinfo {volume} {9}},\
  \bibinfo {pages} {209--222} (\bibinfo {year}
  {2001}{\natexlab{b}})}\BibitemShut {NoStop}%
\end{thebibliography}
\end{document}